\documentclass[twocolumn, amsmath,amssymb, aps, floatfix, nofootinbib]{revtex4-2}

\usepackage{graphicx}% Include figure files
\usepackage{dcolumn}% Align table columns on decimal point
\usepackage{bm}% bold math
\usepackage{hyperref}% add hypertext capabilities
\usepackage{nicefrac}% nice tilted fractions
\usepackage{comment}
\hypersetup{
	colorlinks   = true, %Colours links instead of ugly boxes
	urlcolor     = blue, %Colour for external hyperlinks
	linkcolor    = blue, %Colour of internal links
	citecolor   = blue %Colour of citations, could be ``red''
}
\usepackage{epstopdf}
\usepackage{xcolor}
\usepackage[normalem]{ulem}

\begin{document}

\title{An entanglement-enhanced atomic gravimeter}

\author{Christophe Cassens$^{1,2}$}\email[Author to whom correspondence should be addressed: ]{Christophe.Cassens@dlr.de}
\author{Bernd Meyer-Hoppe$^{1,2}$}
\author{Ernst Rasel$^{1}$}
\author{Carsten Klempt$^{1,2}$}

\affiliation{$^1$Leibniz Universit\"at Hannover, Institut für Quantenoptik, Welfengarten 1, D-30167 Hannover, Germany  \\
$^2$Deutsches Zentrum f\"ur Luft- und Raumfahrt e.V. (DLR), Institut f\"ur Satellitengeod\"asie und Inertialsensorik, Callinstraße 30b, D-30167 Hannover, Germany
}

\date{\today}

\begin{abstract}
%Atom interferometers enable an absolute measurement of the gravitational acceleration with unprecedented precision.
%However, their resolution is fundamentally restricted by quantum fluctuations.
%Outperforming the standard quantum limit requires entangled or squeezed atoms.
%While entanglement-enhanced sensitivities were demonstrated in internal-state measurements and were recently extended to momentum-state atom interferometers, their application for retrieving an inertial or gravitational signal is still pending.
%Here, we present an atomic gravimeter with a sensitivity of $-1.7^{+0.4}_{-0.5}$~dB beyond the standard quantum limit.
%Our entangled atom source is based on ultracold atomic Bose-Einstein condensates whose velocity distribution is further narrowed by delta-kick collimation.
%The low expansion velocities and the entanglement scheme are directly applicable to enhance the resolution of large-scale atom interferometers beyond the classical limits.
Interferometers based on ultra-cold atoms enable an absolute measurement of inertial forces with unprecedented precision. However, their resolution is fundamentally restricted by quantum fluctuations.
Improved resolutions with entangled or squeezed atoms were demonstrated in internal-state measurements for thermal and quantum-degenerate atoms and, recently, for momentum-state interferometers with laser-cooled atoms.
Here, we present a gravimeter based on Bose-Einstein condensates with a sensitivity of $-1.7^{+0.4}_{-0.5}$~dB beyond the standard quantum limit. 
Interferometry with Bose-Einstein condensates combined with delta-kick collimation minimizes atom loss in and improves scalability of the interferometer to very-long baseline atom interferometers.

\end{abstract}

\maketitle

\begin{figure*}[ht]
\centering
  \includegraphics[width=\textwidth]{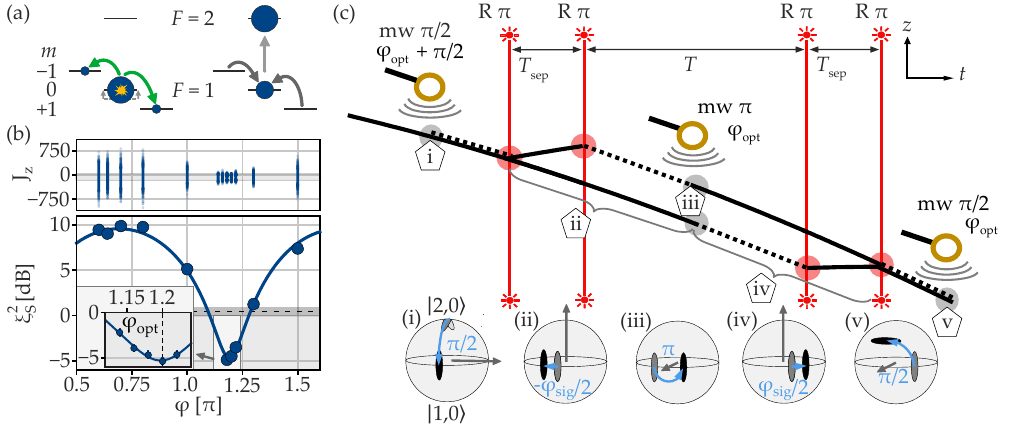}
  \caption{
  The entanglement-enhanced gravimeter.
  (a) Generation of two-mode squeezing by spin-changing collisions (green) via mw dressing, followed by a transfer to clock states via mw (light gray) and rf (dark gray) pulses.
  (b) Spin-noise tomography of the interferometer input state.
  The upper graph shows the normalized population in $|2,0\rangle$ depending on the scanned mw phase $\varphi$ with blue dots as individual measurements.
  In the bottom graph, the variance of these data points is compared to the SQL yielding a spin-squeezing parameter $\xi_\text{S}^2$. 
  The inset shows a detailed measurement around the minimum that represents the optimal squeezing angle $\varphi_\text{opt}$ (dashed line).
  The solid blue line presents a sinusoidal fit to the data, the dashed black line and gray area the experimental coherent state with $0.5^{+0.4}_{-0.5}\,$dB and the hatched area the spin-squeezed regime with sub-SQL fluctuations. 
  Error bars of squeezing parameters are the statistical $\pm1$ standard error \cite{hamley_spinnematic_2012}, but are smaller than the marker size.
  (c) Interferometric sequence in a space-time diagram (top) and in Bloch-sphere representation (bottom).
  Dashed lines indicate the hyperfine level $|1,0\rangle$ and solid lines $|2,0\rangle$.
  Hyperfine levels are changed by mw and Raman (R) pulses.
  Raman pulses also induce a state-dependent change in momentum mode.
  For the Bloch spheres, blue arrows indicate rotations and gray arrows the respective rotation axes.
  The north pole corresponds to $|2,0\rangle$ and the south pole to $|1,0\rangle$.
  The squeezed input state is rotated into the phase-squeezed direction (i) and senses a phase $\varphi_\text{sig}$ (ii-iv) that is finally mapped onto a population imbalance for readout (v).
  }\label{fig:sequenceScheme} 
\end{figure*}

Atom interferometers, in particular light-pulse interferometers, are employed for sensing gravitational fields, with applications for gravimetry~\cite{mcguirk_large_2000}, gradiometry~\cite{asenbaum_phase_2017,chiow_gravitygradient_2017,biedermann_testing_2015,mcguirk_sensitive_2002}, tests of general relativity~\cite{dimopoulos_testing_2007,dimopoulos_general_2008,ufrecht_atominterferometric_2020} and the detection of gravitational waves~\cite{abend_terrestrial_2023,abe_matterwave_2021,badurina_aion_2020,canuel_elgar_2020,tino_sage_2019,canuel_exploring_2018,hogan_atominterferometric_2016,hogan_atomic_2011,dimopoulos_atomic_2008}.
The resolution of the gravity signal is ideally bounded by the standard quantum limit (SQL) that scales with the square root of the atom number. 
Increasing the flux of ultracold atoms is a challenge and moreover, quantum density fluctuations eventually limit the achievable resolution~\cite{feldmann_optimal_2023}.
These limits can be overcome by operating the interferometers with squeezed atomic input states, where entanglement between the atoms enables a suppression of these fundamental signal fluctuations.

Squeezing-enhanced sensitivities were demonstrated in a wide variety of systems~\cite{pezze_quantum_2018}, but mainly in internal degrees of freedom that do not couple to inertial forces.
Entanglement of momentum modes was generated with colliding atoms~\cite{bonneau_tunable_2013, dussarrat_twoparticle_2017, bucker_Twinatom_2011} and in our previous work using atomic Bose-Einstein condensates (BECs)~\cite{anders_momentum_2021}.
Proof-of-principle demonstrations of spin-squeezed Mach-Zehnder interferometers are so far based on laser-cooled atoms ~\cite{malia_distributed_2022, greve_entanglementenhanced_2022}.
%\del{The retrieval of a gravitational signal was not yet reported and the squeezing concepts with optical cavities and use of non-condensed atoms prevent a straight-forward application for high-precision atom interferometry with longer free-fall times and large momentum transfer.
%An operation with \rep{Bose-Einstein condensates}{BECs} is desirable~\cite{szigetti_high_precision,corgier_delta_kick}, as they fulfill the stringent requirements on expansion velocities and spatial-mode control that are set by many high-precision applications.}
The retrieval of a gravitational signal was not yet reported.
%Moreover, high-precision gravimetry scenarios desire long free-fall times and large momentum transfer.
Moreover, high-precision gravimetry scenarios benefit from Bose-Einstein condensed atomic ensembles, because the obtainable narrower position and velocity distributions~\cite{deppner_CollectiveMode_2021} suppress systematic uncertainties~\cite{feldmann_optimal_2023}.
In addition, the small cloud size and velocity width enable a large momentum transfer despite wavefront distortions and the stringent velocity selectivity, respectively~\cite{chiow_102hhbark_2011,debs_Coldatom_2011,gebbe_Twinlattice_2021}. 
A further application of BEC interferometry could be gravimetry at small distances, where small cloud sizes and low densities favor an entanglement-enhancement.
%It is therefore desirable to demonstrate a squeezing concept for gravimetry that is based on BECs~\cite{szigetti_high_precision,corgier_delta_kick}.
The development of a BEC-based squeezing concept for gravimetry is therefore highly desirable~\cite{szigetti_high_precision,corgier_delta_kick}.
%BECs are in principle compatible with long interrogation times due to high contrast and large coherence [cite: Role of source coherence in atom interferometry
%Kyle S. Hardman, Carlos C. N. Kuhn, Gordon D. McDonald, John E. Debs, Shayne Bennetts, John D. Close, and Nicholas P. Robins Phys. Rev. A 89, 023626].
%They also offer beneficial spatial-mode control and thus the possibility of large momentum transfer because of narrow momentum distribution and low expansion velocities [cite: 102hbark Large Area Atom Interferometers, Sheng-wey Chiow, Tim Kovachy, Hui-Chun Chien, and Mark A. Kasevich, Phys. Rev. Lett. 107, 130403 AND/OR Cold-atom gravimetry with a Bose-Einstein condensate, J. E. Debs, P. A. Altin, T. H. Barter, D. Döring, G. R. Dennis, G. McDonald, R. P. Anderson, J. D. Close, and N. P. Robins, Phys. Rev. A 84, 033610 AND/OR something else].

Here we report the application of squeezed states in rubidium BECs to measure the gravitational acceleration with a sensitivity beyond the SQL. 
Two-mode squeezing is generated by spin-changing collisions, and transferred to single-mode squeezing on the magnetic-field-insensitive clock transition.
Microwave (mw) and Raman-laser pulses are combined to form a gravity-sensitive atom interferometer.
The input state with $-5.4^{+0.4}_{-0.5}$~dB spin squeezing enables an interferometer operation with a sensitivity of $-3.9^{+0.6}_{-0.7}$~dB below the experimentally recorded coherent-state reference and $-1.7^{+0.4}_{-0.5}$~dB below the theoretical SQL.
An alternating operation of two interferometer sequences with different interrogation times yields an absolute measurement of the gravitational acceleration.
Our concept can be implemented in existing large-scale BEC-based atom interferometers with small integration efforts.

We initially create BECs of $6\cdot 10^3$ $^{87}$Rb atoms in a crossed-beam optical dipole trap with trapping frequencies $2\pi \times \{150, 160, 220\}\,$Hz.
The atoms are prepared in spin level $|F,m \rangle=|1,0 \rangle$ in a homogeneous magnetic field of $90\,$\textmu{}T that is actively stabilized within $\pm 7\,$nT and oriented in parallel to the Earth's gravitational field.
Due to the quadratic Zeeman shift, the creation of pairs of atoms in $|1,\pm1 \rangle$ caused by spin-changing collisions~\cite{hamley_spinnematic_2012,peise_satisfying_2015} is in principle prevented.
We activate this spin dynamics by dressing the clock transition ($|1,0\rangle \leftrightarrow |2,0\rangle$) with a blue-detuned mw field for $50\,$ms, thus compensating the quadratic Zeeman shift.
This populates the levels $|1,\pm1 \rangle$ with a two-mode squeezed vacuum state (Fig.~\ref{fig:sequenceScheme} (a) left) according to the Hamiltonian~\cite{kruse_improvement_2016}
\begin{equation}
    \hat{H}=\hat{H}_a - \hat{H}_s
\end{equation}
with 
\begin{equation}
    \hat{H}_{s/a}=\frac{\Omega}{2}\left( \hat{a}_{s/a} \hat{a}_{s/a} + \hat{a}_{s/a}^{\dagger} \hat{a}_{s/a}^{\dagger} \right) \text{.}
\end{equation}
%Both contributions $\hat{H}_s$ and $\hat{H}_a$ create and annihilate pairs of atoms in a single mode, i.e. the symmetric ($s$) and antisymmetric ($a$) superposition, respectively, with annihilation (creation) operators
The Hamiltonian contains pairs of the operators
\begin{equation}
    \hat{a}_{s/a}^{(\dagger)}=\frac{1}{\sqrt{2}} \left( \hat{a}_{+1}^{(\dagger)}\pm\hat{a}_{-1}^{(\dagger)} \right)
\end{equation}
which create and annihilate pairs of atoms in the symmetric ($s$) and antisymmetric ($a$) superposition of the levels $|1, \pm 1\rangle$.
We measure a spin dynamics interaction strength of $\Omega=h\times 3.77\,$Hz.
The experimental sequence continues by switching off the dipole trap, such that density-dependent interactions cease. 
After $1$\,ms of free fall, the dipole trap is turned on again for $350\,$\textmu s in order to slow down the expansion of the cloud~\cite{ammann_delta_1997,anders_momentum_2021}. 
Up to this point, the spin-squeezed state is magnetic-field sensitive to first order.
To transfer the squeezed vacuum in $|1,\pm 1 \rangle$ to a magnetically-insensitive clock state, the large amount of atoms in $|1,0 \rangle$ is first transferred to $|2,0 \rangle$ by a mw $\pi$-pulse.
Atoms in $|1,\pm 1 \rangle$ are then transferred to $|1,0 \rangle$ by a $\sigma^{-}$-polarized radio-frequency (rf) $\pi$-pulse with phase $\phi_\text{rf}$ (Fig.~\ref{fig:sequenceScheme}(a) right) that leaves the atoms in $F=2$ unaffected~\cite{kunkel_simultaneous_2019}.
The rf pulse couples only to the symmetric superposition and thus transfers a single-mode squeezed state to $|1,0 \rangle$, containing on average $1.9$~atoms~\cite{supp}.
The few remaining atoms in the levels $|1,\pm 1 \rangle$ do not contribute to the interferometer sequence.
The combination of the single-mode squeezed vacuum state in $|1,0\rangle$ and the large number of atoms in $|2,0\rangle$ constitutes a spin-squeezed state as a basis for the entanglement-enhanced gravimeter described below.  
%After the atoms are now in a squeezed state distributed among the clock states $|1,0\rangle$ and $|2,0\rangle$, we perform an interferometric sequence to measure gravitational acceleration that is described in detail later on.
%For the determination of atom number distributions after the sequence, 
For the final detection of the interferometer output, we separate Zeeman levels with different $m$ by a strong magnetic-field gradient and employ absorption detection~\cite{supp}.

Before the description of the gravimeter, we present a characterization of the single-mode spin-squeezed state in a spin-noise-tomography measurement~\cite{pezze_quantum_2018} based on mw pulses on the clock transition with Rabi frequency $\Omega_\text{mw}=2 \pi \times 20.9\,$kHz and duration $\tau$.
The population imbalance 
\begin{align}
    J_\text{z} = \frac{1}{2}\left(N_{F=2}-N_{F=1}\right)
\end{align} 
%\del{is recorded after the spin-echo sequence with the mw pulses only.}
is recorded after a spin-echo sequence.
This sequence consists of three mw pulses of type $(\tfrac{\pi}{2})_{\varphi+\pi/2}, (\pi)_{\varphi}, (\tfrac{\pi}{2})_{\varphi}$ 
%where for a rotation $\theta_{\phi}$ the parameter $\theta$ represents the rotation angle and $\phi$ the azimuthal angle of the rotation axis~\cite{Dunning_composite_2014}.
where $\theta_{\phi}$ specifies the mw pulse by $\theta=\Omega_\text{mw} \tau$ and the adjustable mw phase $\phi$.
%~\cite{dunning_Composite_2014}}
%The parameter $\phi$ can be experimentally adjusted by the phase of the mw pulses.
%The interferometer output after the echo sequence is insensitive to microwave frequency noise and clock signals.}
%The tomography is performed with the spin echo sequence, as it forms the basis of our gravimeter sequence, as detailed below.
Fig.~\ref{fig:sequenceScheme}(b) shows fluctuations of $J_\text{z}$ as a function of the scanned mw phase $\varphi$ with respect to the rf phase $\phi_\text{rf}$.
%\blue{This common phase matters because it is in relation to the orientation of the squeezing ellipse.}
From the variance of the measured data at each phase, we obtain a spin-squeezing parameter~\cite{pezze_quantum_2018} of 
\begin{align}
    \xi_S^2=4\frac{\text{Var}(J_\text{z})}{N}\hat{=}-5.4^{+0.4}_{-0.5}\,\text{dB}
\end{align}
at a mw phase of $\varphi_\text{opt}=1.2\,\pi$.
The corresponding antisqueezing amounts to $9.9^{+0.4}_{-0.5}\,$dB at $\varphi_\text{opt} + \frac{\pi}{2}$.
This spin-squeezed state is insensitive to magnetic field fluctuations to first order and subsequently employed to decrease the quantum noise in an inertially sensitive interferometer sequence.

The interferometric measurement (Fig.~\ref{fig:sequenceScheme} (c)) is now a combination of the spin-echo sequence from above (i, iii, v) with four Raman-laser pulses that form a gravity-sensitive Mach-Zehnder interferometer.
It starts by applying a mw $\frac{\pi}{2}$-pulse with a phase $\varphi_{\text{opt}} + \frac{\pi}{2}$.
In this orientation, the state features a minimal uncertainty of the phase between the two clock states (i). 
After $1.9\,$ms, a Raman $\pi$-pulse driving the transition $|1,0; p=0\rangle \rightarrow |2,0; p=\hbar k_\text{eff}\rangle$ with  $98.1(7)\,\%$ efficiency renders the interferometer sensitive to acceleration (ii).
The Raman pulse transfers two-photon momenta $\hbar k_\text{eff}=1.18\frac{\text{cm}}{\text{s}}\times m_\text{Rb-87}$ with $m_\text{Rb-87}$ being the atomic mass, leading to a spatial delocalization of the two momentum modes.
A description and characterization of the Raman laser system can be found in the Supplementary Material~\cite{supp}.
The $\tau_\text{R} = 60\,$\textmu s long Raman pulse is Blackman-shaped~\cite{meyer-hoppe_dynamical_2023} to suppress the unwanted transition $|2,0; p=0\hbar k_\text{eff}\rangle$ to $|1,0; p=-\hbar k_\text{eff}\rangle$, which is only $2\pi\times30\,$kHz detuned.
The two clouds separate for $T_\text{sep} = 77\,$\textmu s, before a second Raman $\pi$-pulse decelerates the upper arm of the interferometer by driving the same transition. 
While the two clouds fall in the same momentum mode for a time $T$, the internal states are inverted by a resonant mw $\pi$-pulse to echo the spin evolution and suppress common noise like differential AC-Stark shifts, mw and Raman phase noise, and systematic mw frequency offsets (iii).
After the inverting echo pulse, the two arms acquire an additional gravitational phase shift, now with opposite sign (iv), such that it is not canceled by the echo sequence.
The clouds are reunited by performing the identical Raman processes on the lower arm of the interferometer. 
$1.9\,$ms after the mw $\pi$-pulse, the imprinted inertial phase with squeezed quantum noise is mapped onto the population imbalance $J_{z}$ by a mw $\frac{\pi}{2}$-pulse with phase $\varphi_{\text{opt}}$ (v). 
The time between the final Raman and the  closing mw pulse allows
additional parasitic interferometer paths, that arise from incomplete Raman transfers, to 
detach from the main paths.

The frequency difference of the Raman laser beams is switched between the pulses according to a frequency chirp-rate which is varied around the value $\alpha=9.8126\,\text{m}/\text{s}^2\times k_\text{eff}\,$.
This frequency chirp counteracts the gravitational phase $g\times k_\text{eff}$ perceived by the freely falling atoms, yielding a recorded phase signal of
\begin{align}
    \varphi_\text{sig} = \left(g-\frac{\alpha}{k_\text{eff}}\right)S(T, T_\text{sep}, \tau_\text{R} ).
    \label{eq:gravimeterPhase}
\end{align}

\begin{figure}[t h!]
\centering
  \includegraphics[width=\columnwidth]{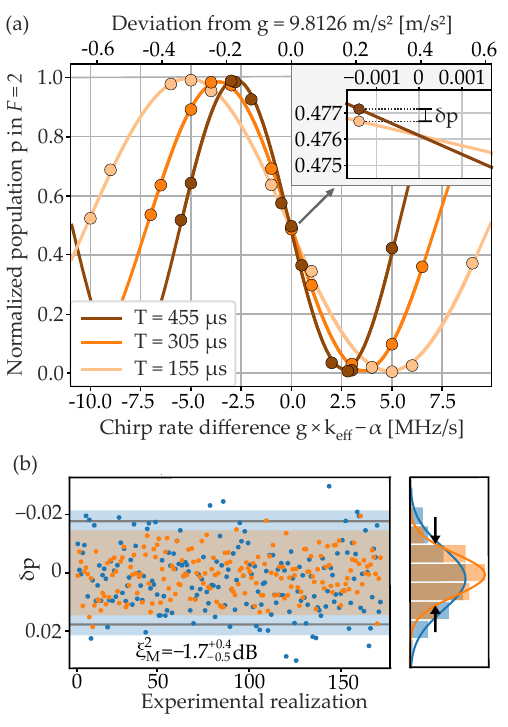}
    \caption{
  Contrast and mid-fringe measurement.
  (a) Interferometer fringes for three different times $T$ of the squeezed gravimeter depicted in Fig.~\ref{fig:sequenceScheme}(c) over chirp rate $\alpha$.
  From the sinusoidal fits, we infer a contrast of $98.0(1.4)\,\%$. 
  All interferometer fringes intercept for the chirp rate which compensates the gravitational acceleration of the atoms in accordance with Eq.~\ref{eq:gravimeterPhase}. 
  The standard deviation of the measurements is smaller than the markers.
  The inset zooms into the crossing region from which the local gravitation $g_\text{exp}$ is obtained by a measurement of the difference signal $\delta p$.
  (b) Gravimetry signal $\delta p$ as interferometer output.
  The gravimeter signal of the squeezing-enhanced operation (orange) has reduced fluctuations compared to the operation with a coherent state (blue).
  Experimentally, we achieve a metrologically relevant reduction of the variance of $-3.9^{+0.6}_{-0.7}\,$dB compared to the experimentally recorded coherent input state and $\xi_\text{M}^2=-1.7^{+0.4}_{-0.5}\,$dB compared to the theoretical SQL.
  Shaded areas indicate $\pm2$ standard deviations of the respective data, grey lines $\pm2$ standard deviations corresponding to the SQL.
  }
  \label{fig:chirprate}
\end{figure}

Here, g is the gravitational acceleration of the atoms and $S$ is a scale factor depending on the interferometer geometry and the Raman pulse duration and shape~\cite{fang_bess_improving_2018}. 

If the chirp rate is set to exactly cancel the gravitational shift, the imprinted inertial phase vanishes according to Eq.~\ref{eq:gravimeterPhase} for all scaling factors. 
By choosing a phase difference of $\frac{\pi}{2}$ between the opening and closing mw $\frac{\pi}{2}$-pulses,  the accumulated phase maps to a measurement of $J_z=0$ (mid-fringe position).
Fig.~\ref{fig:chirprate}(a) shows the normalized population in $F=2$ for three different durations $T$. 
From the interception of the curves, an approximate value for the compensating chirp rate and a corresponding working range can be extracted.

The determination of the gravitational acceleration and the entanglement-enhanced sensitivity is performed by an alternating measurement of the normalized population in $F=2$ for two different durations $T_1=455\,$\textmu s and $T_2=155\,$\textmu s at the previously determined chirp-rate.
This alternating operation suppresses the influence of drifts of the Raman pulse efficiencies that are slow with respect to the cycle time of $52\,$s.
Since the scale factors and the corresponding slopes at this point differ, an experimental value for the gravitational acceleration $g_\text{exp}$ can be obtained from the difference 
\begin{align}
    \delta  p = p(T_1) - p(T_2) = \frac{N_\text{F=2}(T_1)}{N(T_1)}-\frac{N_\text{F=2}(T_2)}{N(T_2)}
\end{align} 
of the normalized populations of $|F=2\rangle$ (see Fig.~\ref{fig:chirprate} (a) inset) according to 
\begin{align}
    \label{eq:gFromPopDeiff}
    g_\text{exp} = \frac{2}{C}\frac{\delta p}{S(T_1)-S(T_2)} + \frac{\alpha}{k_\text{eff}} 
\end{align}
with contrast $C=98.0(1.4)\,\%$ obtained from the full fringes in Fig.~\ref{fig:chirprate}(a) and respective scale factors $S(T_1)=-1.42(1)\,\frac{\text{s}}{\text{m}^2}$ and $S(T_2)=-0.767(3)\,\frac{\text{s}}{\text{m}^2}$. 
The measured value of $g_\text{exp}=9.8118(16)\,\frac{\text{m}}{\text{s}^2}$ agrees with the local gravitational acceleration of $9.812\,637\,196\,(88)$~m/s$^2$~\cite{hartwig_analyse_2013} within the bounds of one standard error.
Note that the error of the scale factors contributes much less to the uncertainty in $g_\text{exp}$ than the statistical uncertainty. 
Furthermore, the scale factors inferred from the fringe measurements are reproduced by the theoretical calculation~\cite{supp}.

We evaluate the metrological improvement by comparing the recorded measurement fluctuations with the optimal result from an ideal unentangled coherent state. 
We obtain a metrological squeezing factor 
\begin{align}
\xi_\text{M}^2= \frac{4}{C^2}\frac{\text{Var}(J_\text{z}(T_1)-J_\text{z}(T_2))}{N(T_2)+N(T_1)}\hat{=}-1.7^{+0.4}_{-0.5}\,\text{dB,}
\end{align}
where $(N(T_2)+N(T_1))/4$ is the SQL of the difference of two independent measurements with their respective atom numbers $N(T_2)$ and $N(T_1)$.
This improvement proves the entanglement-enhanced measurement of the gravitational acceleration and constitutes the main result of our work.
The corresponding data is shown in Fig.~\ref{fig:chirprate}(b) in comparison to a coherent state realization.

\begin{figure}[t]
\centering
  \includegraphics[width=\columnwidth]{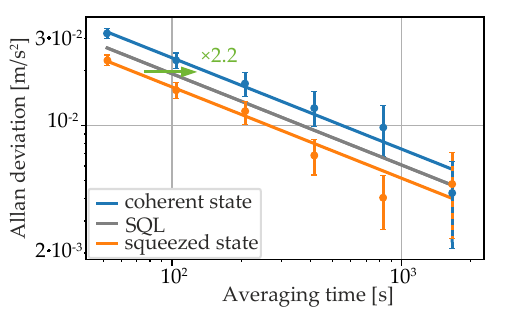} 
    \caption{
  Allan deviation of the gravimeter sequence using coherent states (blue) and squeezed states (orange).
  %The blue data points belong to the gravimeter sequence with a squeezed input state as described above to which $0.081/\sqrt{\tau}\,\frac{\text{m}}{\text{s}^2\sqrt{\text{Hz}}}$ is fitted.
  %$0.119/\sqrt{\tau}\,\frac{\text{m}}{\text{s}^2\sqrt{\text{Hz}}}$ is fitted to the same gravimeter protocol employing a coherent input state (orange). 
  The squeezing-enhanced sequence reaches any instability $1.4$-times faster than the theoretical optimum at the SQL (gray line) and $2.2$-times faster than the experimentally recorded equivalent with coherent states. 
  The technical noise sources for both sequences were the same. 
  Error bars are the statistical $\pm1$ standard error \cite{riley_handbook_2008}.
  }
  \label{fig:allandev}
\end{figure}

We further analyze the temporal behavior throughout the measurement runs by calculating the Allan deviations \cite{riley_handbook_2008} of $\delta p$ for a coherent and a squeezed input state shown in Fig.~\ref{fig:allandev}.
The coherent-state case is realized by omitting the squeezing generation section in the original experimental sequence. 
The squeezed-state results outperform the coherent-state equivalent over the whole range of averaging times.
From a fit to the first $800\,$s of averaging time, we conclude that the squeezed-state signal averages down $2.2$-times faster than the coherent-state signal and $1.4$-times faster than the SQL for the same number of employed atoms. 

%To determine metrological improvement the Wineland parameter is used \cite{wineland_squeezed_1994}.
%Just as the spin-squeezing parameter $\xi_\textr{S}^2$ it takes into account the reduced variance of measured $\delta$ (see Fig~\ref{fig:sequenceScheme}(d)) compared to the SQL (here given by $2\times\frac{N}{4}$) and additionally also the reduction of contrast $C$ of the gravimeter signal:

%\begin{align}
%    \xi_\text{R}^2=W=\frac{2\text{var}(\delta)}{NC^2}
%\end{align}

%The contrast is determined from the interferometer fringes in Fig.~\ref{fig:chirprate}.
%The obtained Wineland parameter is $\xi_\text{R}^2=W = -1.51^{+0.32}_{-0.34}\,$dB \comm{@Carsten: Do we want to write W or $\xi_\text{R}^2$?}.
%As another measure of improvement in metrological performance due to squeezing we compare the overlapping Allan deviations \cite{riley_handbook_2008} of measured $\delta$ for a coherent and a squeezed input state shown in Fig.~\ref{fig:allandev}.
%In case of the conventional coherent input state the gravimeter sequence outlined above is repeated skipping the squeezing generation.
%It is clearly visible that the sequence employing entangled states outperforms the classical sequence over the whole range of averaging times.
%From a fit to the first $800\,$s of averaging time we conclude that the entanglement improved gravimeter averages down to any given instability $2.2$-times faster than it would without squeezing the input state and $1.4$-times faster than the solely SQL-limited sequence for the same number of employed atoms would.

In summary, we have presented a concept for enhancing atomic gravimeters beyond the SQL. 
Our demonstration involves all components for a large-scale implementation aiming for highest sensitivities. 
The squeezing method can be implemented in existing BEC-based atomic sources and can be scaled to large atom numbers due to the utilization of vacuum squeezing. 
The measured antisqueezing is lower compared to other squeezing protocols like cavity-QND and helps to maintain a large contrast and dynamical range in the interferometry signal. 
The implementation with BECs provides low expansion velocities and exquisite control of the spatial mode to suppress systematic effects due to laser wavefront curvature and distortion or Coriolis force~\cite{peters_measurement_1999}. 
We have shown that the squeezing is compatible with a further delta-kick reduction of the expansion velocity as required for long interrogation times. 
The squeezing angle can be freely adjusted and enables the anticipation and suppression of density-dependent quantum fluctuations~\cite{feldmann_optimal_2023}.

The observed reduction of the squeezing due to the interferometer sequence stems solely from technical noise of the Raman laser system which constitutes an independent task for reaching highest sensitivities. 
Compared to conventional Mach-Zehnder interferometers, our concept features equal spin states during separation and recombination, suppressing the sensitivity to light shifts, mw shifts and magnetic field noise. 

Our method recommends itself for differential measurements, as performed in gradiometry, tests of the universality of free fall, or gravitational wave detection.
In such configurations, technical noise, e.g. induced by vibrations, is common mode and cancels, enabling the exploitation of entanglement enhancement.
We envision the application of entanglement-enhanced interferometry at much increased interrogation times, either in large-scale fountains like the Very-Long Baseline Atom Interferometer~\cite{lezeik_understanding_2022} or in microgravity environments like the Einstein Elevator~\cite{lotz_einsteinelevator_2017}. 
The latter is currently pioneered by the INTENTAS~\cite{anton_INTENTAS_2024} project which aims at demonstrating an entanglement enhancement for future space-borne high-precision atom interferometers.
Further applications include measurements of fundamental constants \cite{Morel_Determination_2020, Atom_Fixler_2007, Rosi_Precision_2014} as well as tests of classicalization~\cite{Schrinski_testing_2023}.

\section*{Data availability}
The data that support the findings of this study are available at Ref.~\cite{CassensRepository2025}.

\section*{Acknowledgments}
We thank Fabian Anders, Th\'eo Sanchez, Dennis Schlippert, Christian Schubert, Polina Feldmann, Luis Santos, Jens Kruse, Naceur Gaaloul for inspiring discussions.
We acknowledge financial support from the Deutsche Forschungsgemeinschaft (DFG, German Research Foundation) -- Project-ID 274200144 -- SFB 1227 DQ-mat within the projects A02 and B01 and under Germany’s Excellence Strategy -- EXC-2123 QuantumFrontiers -- 390837967.
B.M.-H. acknowledges support from the Hannover School for Nanotechnology (HSN).
We acknowledge funding by the SQUEIS project -- grant No 499225223 -- within the QuantERA II Programme.
\section*{Author contributions}

\section*{Competing interests}
The authors declare no competing interests.

%\printbibliography
%\bibliographystyle{unsrturl}
%\bibliography{gravibib}

%\bibliographystyle{apsrev4-2}
%\bibliography{gravibib.bib}

\begin{thebibliography}{53}%
\makeatletter
\providecommand \@ifxundefined [1]{%
 \@ifx{#1\undefined}
}%
\providecommand \@ifnum [1]{%
 \ifnum #1\expandafter \@firstoftwo
 \else \expandafter \@secondoftwo
 \fi
}%
\providecommand \@ifx [1]{%
 \ifx #1\expandafter \@firstoftwo
 \else \expandafter \@secondoftwo
 \fi
}%
\providecommand \natexlab [1]{#1}%
\providecommand \enquote  [1]{``#1''}%
\providecommand \bibnamefont  [1]{#1}%
\providecommand \bibfnamefont [1]{#1}%
\providecommand \citenamefont [1]{#1}%
\providecommand \href@noop [0]{\@secondoftwo}%
\providecommand \href [0]{\begingroup \@sanitize@url \@href}%
\providecommand \@href[1]{\@@startlink{#1}\@@href}%
\providecommand \@@href[1]{\endgroup#1\@@endlink}%
\providecommand \@sanitize@url [0]{\catcode `\\12\catcode `\$12\catcode
  `\&12\catcode `\#12\catcode `\^12\catcode `\_12\catcode `\%12\relax}%
\providecommand \@@startlink[1]{}%
\providecommand \@@endlink[0]{}%
\providecommand \url  [0]{\begingroup\@sanitize@url \@url }%
\providecommand \@url [1]{\endgroup\@href {#1}{\urlprefix }}%
\providecommand \urlprefix  [0]{URL }%
\providecommand \Eprint [0]{\href }%
\providecommand \doibase [0]{https://doi.org/}%
\providecommand \selectlanguage [0]{\@gobble}%
\providecommand \bibinfo  [0]{\@secondoftwo}%
\providecommand \bibfield  [0]{\@secondoftwo}%
\providecommand \translation [1]{[#1]}%
\providecommand \BibitemOpen [0]{}%
\providecommand \bibitemStop [0]{}%
\providecommand \bibitemNoStop [0]{.\EOS\space}%
\providecommand \EOS [0]{\spacefactor3000\relax}%
\providecommand \BibitemShut  [1]{\csname bibitem#1\endcsname}%
\let\auto@bib@innerbib\@empty
%</preamble>
\bibitem [{\citenamefont {McGuirk}\ \emph {et~al.}(2000)\citenamefont
  {McGuirk}, \citenamefont {Snadden},\ and\ \citenamefont
  {Kasevich}}]{mcguirk_large_2000}%
  \BibitemOpen
  \bibfield  {author} {\bibinfo {author} {\bibfnamefont {J.~M.}\ \bibnamefont
  {McGuirk}}, \bibinfo {author} {\bibfnamefont {M.~J.}\ \bibnamefont
  {Snadden}},\ and\ \bibinfo {author} {\bibfnamefont {M.~A.}\ \bibnamefont
  {Kasevich}},\ } 
  \bibfield {title} {\bibinfo {title} \textit{Large Area Light-Pulse Atom Interferometry},\ } \href {https://doi.org/10.1103/PhysRevLett.85.4498} {\bibfield
   {journal} {\bibinfo  {journal} {Phys. Rev. Lett.}\ }\textbf {\bibinfo
  {volume} {85}},\ \bibinfo {pages} {4498} (\bibinfo {year}
  {2000})}\BibitemShut {NoStop}%
\bibitem [{\citenamefont {Asenbaum}\ \emph {et~al.}(2017)\citenamefont 
  {Asenbaum}, \citenamefont {Overstreet}, \citenamefont {Kovachy},
  \citenamefont {Brown}, \citenamefont {Hogan},\ and\ \citenamefont
  {Kasevich}}]{asenbaum_phase_2017}%
  \BibitemOpen
  \bibfield  {author} {\bibinfo {author} {\bibfnamefont {P.}~\bibnamefont
  {Asenbaum}}, \bibinfo {author} {\bibfnamefont {C.}~\bibnamefont
  {Overstreet}}, \bibinfo {author} {\bibfnamefont {T.}~\bibnamefont {Kovachy}},
  \bibinfo {author} {\bibfnamefont {D.~D.}\ \bibnamefont {Brown}}, \bibinfo
  {author} {\bibfnamefont {J.~M.}\ \bibnamefont {Hogan}},\ and\ \bibinfo
  {author} {\bibfnamefont {M.~A.}\ \bibnamefont {Kasevich}},\ }
  \bibfield {title} {\bibinfo {title} \textit{Phase Shift in an Atom Interferometer due to Spacetime Curvature across its Wave Function},\ }\href
  {https://doi.org/10.1103/PhysRevLett.118.183602} {\bibfield  {journal}
  {\bibinfo  {journal} {Phys. Rev. Lett.}\ }\textbf {\bibinfo {volume} {118}},\
  \bibinfo {pages} {183602} (\bibinfo {year} {2017})}\BibitemShut {NoStop}%
\bibitem [{\citenamefont {Chiow}\ \emph {et~al.}(2017)\citenamefont {Chiow},
  \citenamefont {Williams}, \citenamefont {Yu},\ and\ \citenamefont
  {M{\"u}ller}}]{chiow_gravitygradient_2017}%
  \BibitemOpen
  \bibfield  {author} {\bibinfo {author} {\bibfnamefont {S.-w.}\ \bibnamefont
  {Chiow}}, \bibinfo {author} {\bibfnamefont {J.}~\bibnamefont {Williams}},
  \bibinfo {author} {\bibfnamefont {N.}~\bibnamefont {Yu}},\ and\ \bibinfo
  {author} {\bibfnamefont {H.}~\bibnamefont {M{\"u}ller}},\ }
  \bibfield {title} {\bibinfo {title} \textit{Gravity-gradient suppression in spaceborne atomic tests of the equivalence principle},\ }\href
  {https://doi.org/10.1103/PhysRevA.95.021603} {\bibfield  {journal} {\bibinfo
  {journal} {Phys. Rev. A}\ }\textbf {\bibinfo {volume} {95}},\ \bibinfo
  {pages} {021603} (\bibinfo {year} {2017})}\BibitemShut {NoStop}%
\bibitem [{\citenamefont {Biedermann}\ \emph {et~al.}(2015)\citenamefont
  {Biedermann}, \citenamefont {Wu}, \citenamefont {Deslauriers}, \citenamefont
  {Roy}, \citenamefont {Mahadeswaraswamy},\ and\ \citenamefont
  {Kasevich}}]{biedermann_testing_2015}%
  \BibitemOpen
  \bibfield  {author} {\bibinfo {author} {\bibfnamefont {G.~W.}\ \bibnamefont
  {Biedermann}}, \bibinfo {author} {\bibfnamefont {X.}~\bibnamefont {Wu}},
  \bibinfo {author} {\bibfnamefont {L.}~\bibnamefont {Deslauriers}}, \bibinfo
  {author} {\bibfnamefont {S.}~\bibnamefont {Roy}}, \bibinfo {author}
  {\bibfnamefont {C.}~\bibnamefont {Mahadeswaraswamy}},\ and\ \bibinfo {author}
  {\bibfnamefont {M.~A.}\ \bibnamefont {Kasevich}},\ }
  \bibfield {title} {\bibinfo {title} \textit{Testing gravity with cold-atom interferometers},\ }\href
  {https://doi.org/10.1103/PhysRevA.91.033629} {\bibfield  {journal} {\bibinfo
  {journal} {Phys. Rev. A}\ }\textbf {\bibinfo {volume} {91}},\ \bibinfo
  {pages} {033629} (\bibinfo {year} {2015})}\BibitemShut {NoStop}%
\bibitem [{\citenamefont {McGuirk}\ \emph {et~al.}(2002)\citenamefont
  {McGuirk}, \citenamefont {Foster}, \citenamefont {Fixler}, \citenamefont
  {Snadden},\ and\ \citenamefont {Kasevich}}]{mcguirk_sensitive_2002}%
  \BibitemOpen
  \bibfield  {author} {\bibinfo {author} {\bibfnamefont {J.~M.}\ \bibnamefont
  {McGuirk}}, \bibinfo {author} {\bibfnamefont {G.~T.}\ \bibnamefont {Foster}},
  \bibinfo {author} {\bibfnamefont {J.~B.}\ \bibnamefont {Fixler}}, \bibinfo
  {author} {\bibfnamefont {M.~J.}\ \bibnamefont {Snadden}},\ and\ \bibinfo
  {author} {\bibfnamefont {M.~A.}\ \bibnamefont {Kasevich}},\ }
  \bibfield {title} {\bibinfo {title} \textit{Sensitive absolute-gravity gradiometry using atom interferometry},\ }\href
  {https://doi.org/10.1103/PhysRevA.65.033608} {\bibfield  {journal} {\bibinfo
  {journal} {Phys. Rev. A}\ }\textbf {\bibinfo {volume} {65}},\ \bibinfo
  {pages} {033608} (\bibinfo {year} {2002})}\BibitemShut {NoStop}%
\bibitem [{\citenamefont {Dimopoulos}\ \emph {et~al.}(2007)\citenamefont
  {Dimopoulos}, \citenamefont {Graham}, \citenamefont {Hogan},\ and\
  \citenamefont {Kasevich}}]{dimopoulos_testing_2007}%
  \BibitemOpen
  \bibfield  {author} {\bibinfo {author} {\bibfnamefont {S.}~\bibnamefont
  {Dimopoulos}}, \bibinfo {author} {\bibfnamefont {P.~W.}\ \bibnamefont
  {Graham}}, \bibinfo {author} {\bibfnamefont {J.~M.}\ \bibnamefont {Hogan}},\
  and\ \bibinfo {author} {\bibfnamefont {M.~A.}\ \bibnamefont {Kasevich}},\
  }
  \bibfield {title} {\bibinfo {title} \textit{Testing General Relativity with Atom Interferometry},\ }\href {https://doi.org/10.1103/PhysRevLett.98.111102} {\bibfield  {journal}
  {\bibinfo  {journal} {Phys. Rev. Lett.}\ }\textbf {\bibinfo {volume} {98}},\
  \bibinfo {pages} {111102} (\bibinfo {year} {2007})}\BibitemShut {NoStop}%
\bibitem [{\citenamefont {Dimopoulos}\ \emph
  {et~al.}(2008{\natexlab{a}})\citenamefont {Dimopoulos}, \citenamefont
  {Graham}, \citenamefont {Hogan},\ and\ \citenamefont
  {Kasevich}}]{dimopoulos_general_2008}%
  \BibitemOpen
  \bibfield  {author} {\bibinfo {author} {\bibfnamefont {S.}~\bibnamefont
  {Dimopoulos}}, \bibinfo {author} {\bibfnamefont {P.~W.}\ \bibnamefont
  {Graham}}, \bibinfo {author} {\bibfnamefont {J.~M.}\ \bibnamefont {Hogan}},\
  and\ \bibinfo {author} {\bibfnamefont {M.~A.}\ \bibnamefont {Kasevich}},\
  }
  \bibfield {title} {\bibinfo {title} \textit{General relativistic effects in atom interferometry},\ }\href {https://doi.org/10.1103/PhysRevD.78.042003} {\bibfield  {journal}
  {\bibinfo  {journal} {Phys. Rev. D}\ }\textbf {\bibinfo {volume} {78}},\
  \bibinfo {pages} {042003} (\bibinfo {year} {2008}{\natexlab{a}})}\BibitemShut
  {NoStop}%
\bibitem [{\citenamefont {Ufrecht}\ \emph {et~al.}(2020)\citenamefont
  {Ufrecht}, \citenamefont {Di~Pumpo}, \citenamefont {Friedrich}, \citenamefont
  {Roura}, \citenamefont {Schubert}, \citenamefont {Schlippert}, \citenamefont
  {Rasel}, \citenamefont {Schleich},\ and\ \citenamefont
  {Giese}}]{ufrecht_atominterferometric_2020}%
  \BibitemOpen
  \bibfield  {author} {\bibinfo {author} {\bibfnamefont {C.}~\bibnamefont
  {Ufrecht}}, \bibinfo {author} {\bibfnamefont {F.}~\bibnamefont {Di~Pumpo}},
  \bibinfo {author} {\bibfnamefont {A.}~\bibnamefont {Friedrich}}, \bibinfo
  {author} {\bibfnamefont {A.}~\bibnamefont {Roura}}, \bibinfo {author}
  {\bibfnamefont {C.}~\bibnamefont {Schubert}}, \bibinfo {author}
  {\bibfnamefont {D.}~\bibnamefont {Schlippert}}, \bibinfo {author}
  {\bibfnamefont {E.~M.}\ \bibnamefont {Rasel}}, \bibinfo {author}
  {\bibfnamefont {W.~P.}\ \bibnamefont {Schleich}},\ and\ \bibinfo {author}
  {\bibfnamefont {E.}~\bibnamefont {Giese}},\ }\href
  {https://doi.org/10.1103/PhysRevResearch.2.043240} {\bibfield  {journal}
  {\bibinfo  {journal} {Phys. Rev. Res.}\ }\textbf {\bibinfo {volume} {2}},\
  \bibinfo {pages} {043240} (\bibinfo {year} {2020})}\BibitemShut {NoStop}%
\bibitem [{\citenamefont {Abend}\ \emph {et~al.}(2023)\citenamefont {Abend},
  \citenamefont {Allard}, \citenamefont {Alonso}, \citenamefont {Antoniadis},
  \citenamefont {Araujo}, \citenamefont {Arduini}, \citenamefont {Arnold},
  \citenamefont {A{\ss}mann}, \citenamefont {Augst}, \citenamefont {Badurina},
  \citenamefont {Balaz}, \citenamefont {Banks}, \citenamefont {Barone},
  \citenamefont {Barsanti}, \citenamefont {Bassi}, \citenamefont {Battelier},
  \citenamefont {Baynham}, \citenamefont {Quentin}, \citenamefont {Belic},
  \citenamefont {Beniwal}, \citenamefont {Bernabeu}, \citenamefont
  {Bertinelli}, \citenamefont {Bertoldi}, \citenamefont {Biswas}, \citenamefont
  {Blas}, \citenamefont {Boegel}, \citenamefont {Bogojevic}, \citenamefont
  {B{\"o}hm}, \citenamefont {B{\"o}hringer}, \citenamefont {Bongs},
  \citenamefont {Bouyer}, \citenamefont {Brand}, \citenamefont {Brimis},
  \citenamefont {Buchmueller}, \citenamefont {Cacciapuoti}, \citenamefont
  {Calatroni}, \citenamefont {Canuel}, \citenamefont {Caprini}, \citenamefont
  {Caramete}, \citenamefont {Caramete}, \citenamefont {Carlesso}, \citenamefont
  {Carlton}, \citenamefont {Casariego}, \citenamefont {Charmandaris},
  \citenamefont {Chen}, \citenamefont {Chiofalo}, \citenamefont {Cimbri},
  \citenamefont {Coleman}, \citenamefont {Constantin}, \citenamefont
  {Contaldi}, \citenamefont {Cui}, \citenamefont {Da~Ros}, \citenamefont
  {Davies}, \citenamefont {Rosendo}, \citenamefont {Deppner}, \citenamefont
  {Derevianko}, \citenamefont {{de Rham}}, \citenamefont {De~Roeck},
  \citenamefont {Derr}, \citenamefont {Di~Pumpo}, \citenamefont {Djordjevic},
  \citenamefont {Dobrich}, \citenamefont {Domokos}, \citenamefont {Dornan},
  \citenamefont {Doser}, \citenamefont {Drougakis}, \citenamefont {Dunningham},
  \citenamefont {Duspayev}, \citenamefont {Easo}, \citenamefont {Eby},
  \citenamefont {Efremov}, \citenamefont {Ekelof}, \citenamefont {Elertas},
  \citenamefont {Ellis}, \citenamefont {Evans}, \citenamefont {Fadeev},
  \citenamefont {Fan{\`i}}, \citenamefont {Fassi}, \citenamefont {Fattori},
  \citenamefont {Fayet}, \citenamefont {Felea}, \citenamefont {Feng},
  \citenamefont {Friedrich}, \citenamefont {Fuchs}, \citenamefont {Gaaloul},
  \citenamefont {Gao}, \citenamefont {Gardner}, \citenamefont {Garraway},
  \citenamefont {Gauguet}, \citenamefont {Gerlach}, \citenamefont {Gersemann},
  \citenamefont {Gibson}, \citenamefont {Giese}, \citenamefont {Giudice},
  \citenamefont {Glasbrenner}, \citenamefont {G{\"u}ndogan}, \citenamefont
  {Haehnelt}, \citenamefont {Hakulinen}, \citenamefont {Hammerer},
  \citenamefont {Han{\i}meli}, \citenamefont {Harte}, \citenamefont {Hawkins},
  \citenamefont {Hees}, \citenamefont {Heise}, \citenamefont {Henderson},
  \citenamefont {Herrmann}, \citenamefont {Hird}, \citenamefont {Hogan},
  \citenamefont {Holst}, \citenamefont {Holynski}, \citenamefont {Hussain},
  \citenamefont {Janson}, \citenamefont {Jegli{\v c}}, \citenamefont {Jelezko},
  \citenamefont {Kagan}, \citenamefont {Kalliokoski}, \citenamefont {Kasevich},
  \citenamefont {Kehagias}, \citenamefont {Kilian}, \citenamefont {Koley},
  \citenamefont {Konrad}, \citenamefont {Kopp}, \citenamefont {Kornakov},
  \citenamefont {Kovachy}, \citenamefont {Krutzik}, \citenamefont {Kumar},
  \citenamefont {Kumar}, \citenamefont {Laemmerzahl}, \citenamefont
  {Landsberg}, \citenamefont {Langlois}, \citenamefont {Lanigan}, \citenamefont
  {Lellouch}, \citenamefont {Leone}, \citenamefont {Lafitte}, \citenamefont
  {Lewicki}, \citenamefont {Leykauf}, \citenamefont {Lezeik}, \citenamefont
  {Lombriser}, \citenamefont {L{\'o}pez}, \citenamefont {Asamar}, \citenamefont
  {Monjaraz}, \citenamefont {Luciano}, \citenamefont {Mohammed}, \citenamefont
  {Maleknejad}, \citenamefont {Markus}, \citenamefont {Marteau}, \citenamefont
  {Massonnet}, \citenamefont {Mazumdar}, \citenamefont {McCabe}, \citenamefont
  {Meister}, \citenamefont {Menu}, \citenamefont {Messineo}, \citenamefont
  {Micalizio}, \citenamefont {Millington}, \citenamefont {Milosevic},
  \citenamefont {Mitchell}, \citenamefont {Montero}, \citenamefont {Morley},
  \citenamefont {M{\"u}ller}, \citenamefont {M{\"u}stecapl{\i}o{\u g}lu},
  \citenamefont {Ni}, \citenamefont {Noller}, \citenamefont {Od{\v z}ak},
  \citenamefont {Oi}, \citenamefont {Omar}, \citenamefont {Pahl}, \citenamefont
  {Paling}, \citenamefont {Pandey}, \citenamefont {Pappas}, \citenamefont
  {Pareek}, \citenamefont {Pasatembou}, \citenamefont {Pelucchi}, \citenamefont
  {dos Santos}, \citenamefont {Piest}, \citenamefont {Pikovski}, \citenamefont
  {Pilaftsis}, \citenamefont {Plunkett}, \citenamefont {Poggiani},
  \citenamefont {Prevedelli}, \citenamefont {Puputti}, \citenamefont {Veettil},
  \citenamefont {Quenby}, \citenamefont {Rafelski}, \citenamefont {Rajendran},
  \citenamefont {Rasel}, \citenamefont {Sfar}, \citenamefont {Reynaud},
  \citenamefont {Richaud}, \citenamefont {Rodzinka}, \citenamefont {Roura},
  \citenamefont {Rudolph}, \citenamefont {Sabulsky}, \citenamefont {Safronova},
  \citenamefont {Santamaria}, \citenamefont {Schilling}, \citenamefont
  {Schkolnik}, \citenamefont {Schleich}, \citenamefont {Schlippert},
  \citenamefont {Schneider}, \citenamefont {Schreck}, \citenamefont {Schubert},
  \citenamefont {Schwersenz}, \citenamefont {Semakin}, \citenamefont
  {Sergijenko}, \citenamefont {Shao}, \citenamefont {Shipsey}, \citenamefont
  {Singh}, \citenamefont {Smerzi}, \citenamefont {Sopuerta}, \citenamefont
  {Spallicci}, \citenamefont {Stefanescu}, \citenamefont {Stergioulas},
  \citenamefont {Str{\"o}hle}, \citenamefont {Struckmann}, \citenamefont
  {Tentindo}, \citenamefont {Throssell}, \citenamefont {Tino}, \citenamefont
  {Tinsley}, \citenamefont {Mircea}, \citenamefont {Tkal{\v c}ec},
  \citenamefont {Tolley}, \citenamefont {Tornatore}, \citenamefont
  {{Torres-Orjuela}}, \citenamefont {Treutlein}, \citenamefont {Trombettoni},
  \citenamefont {Tsai}, \citenamefont {Ufrecht}, \citenamefont {Ulmer},
  \citenamefont {Valuch}, \citenamefont {Vaskonen}, \citenamefont {Aceves},
  \citenamefont {Vitanov}, \citenamefont {Vogt}, \citenamefont {{von
  Klitzing}}, \citenamefont {Vukics}, \citenamefont {Walser}, \citenamefont
  {Wang}, \citenamefont {Warburton}, \citenamefont {{Webber-Date}},
  \citenamefont {Wenzlawski}, \citenamefont {Werner}, \citenamefont {Williams},
  \citenamefont {Windapssinger}, \citenamefont {Wolf}, \citenamefont
  {W{\"o}rner}, \citenamefont {Xuereb}, \citenamefont {Yahia}, \citenamefont
  {Cruzeiro}, \citenamefont {Zarei}, \citenamefont {Zhan}, \citenamefont
  {Zhou}, \citenamefont {Zupan},\ and\ \citenamefont {Zupani{\v
  c}}}]{abend_terrestrial_2023}%
  \BibitemOpen
  \bibfield  {author} {\bibinfo {author} {\bibfnamefont {S.}~\bibnamefont
  {Abend}} et al. }
  \bibfield {title} {\bibinfo {title} \textit{Terrestrial Very-Long-Baseline Atom Interferometry: Workshop Summary},\ }\href
  {https://doi.org/10.48550/arXiv.2310.08183} {\bibfield  {journal} {\bibinfo
  {journal} {arXiv}\ ,\ \bibinfo {pages} {2310.08183}} (\bibinfo {year}
  {2023})}\BibitemShut {NoStop}%
\bibitem [{\citenamefont {Abe}\ \emph {et~al.}(2021)\citenamefont {Abe},
  \citenamefont {Adamson}, \citenamefont {Borcean}, \citenamefont {Bortoletto},
  \citenamefont {Bridges}, \citenamefont {Carman}, \citenamefont
  {Chattopadhyay}, \citenamefont {Coleman}, \citenamefont {Curfman},
  \citenamefont {DeRose}, \citenamefont {Deshpande}, \citenamefont
  {Dimopoulos}, \citenamefont {Foot}, \citenamefont {Frisch}, \citenamefont
  {Garber}, \citenamefont {Geer}, \citenamefont {Gibson}, \citenamefont
  {Glick}, \citenamefont {Graham}, \citenamefont {Hahn}, \citenamefont
  {Harnik}, \citenamefont {Hawkins}, \citenamefont {Hindley}, \citenamefont
  {Hogan}, \citenamefont {Jiang}, \citenamefont {Kasevich}, \citenamefont
  {Kellett}, \citenamefont {Kiburg}, \citenamefont {Kovachy}, \citenamefont
  {Lykken}, \citenamefont {{March-Russell}}, \citenamefont {Mitchell},
  \citenamefont {Murphy}, \citenamefont {Nantel}, \citenamefont {Nobrega},
  \citenamefont {Plunkett}, \citenamefont {Rajendran}, \citenamefont {Rudolph},
  \citenamefont {Sachdeva}, \citenamefont {Safdari}, \citenamefont {Santucci},
  \citenamefont {Schwartzman}, \citenamefont {Shipsey}, \citenamefont {Swan},
  \citenamefont {Valerio}, \citenamefont {Vasonis}, \citenamefont {Wang},\ and\
  \citenamefont {Wilkason}}]{abe_matterwave_2021}%
  \BibitemOpen
  \bibfield  {author} {\bibinfo {author} {\bibfnamefont {M.}~\bibnamefont
  {Abe}}, \bibinfo {author} {\bibfnamefont {P.}~\bibnamefont {Adamson}},
  \bibinfo {author} {\bibfnamefont {M.}~\bibnamefont {Borcean}}, \bibinfo
  {author} {\bibfnamefont {D.}~\bibnamefont {Bortoletto}}, \bibinfo {author}
  {\bibfnamefont {K.}~\bibnamefont {Bridges}}, \bibinfo {author} {\bibfnamefont
  {S.~P.}\ \bibnamefont {Carman}}, \bibinfo {author} {\bibfnamefont
  {S.}~\bibnamefont {Chattopadhyay}}, \bibinfo {author} {\bibfnamefont
  {J.}~\bibnamefont {Coleman}}, \bibinfo {author} {\bibfnamefont {N.~M.}\
  \bibnamefont {Curfman}}, \bibinfo {author} {\bibfnamefont {K.}~\bibnamefont
  {DeRose}} et al.,\ }
  \bibfield {title} {\bibinfo {title} \textit{Matter-wave Atomic Gradiometer Interferometric Sensor (MAGIS-100)},\ }\href
  {https://doi.org/10.1088/2058-9565/abf719} {\bibfield  {journal} {\bibinfo
  {journal} {Quantum Sci. Technol.}\ }\textbf {\bibinfo {volume} {6}},\
  \bibinfo {pages} {044003} (\bibinfo {year} {2021})}\BibitemShut {NoStop}%
\bibitem [{\citenamefont {Badurina}\ \emph {et~al.}(2020)\citenamefont
  {Badurina}, \citenamefont {Bentine}, \citenamefont {Blas}, \citenamefont
  {Bongs}, \citenamefont {Bortoletto}, \citenamefont {Bowcock}, \citenamefont
  {Bridges}, \citenamefont {Bowden}, \citenamefont {Buchmueller}, \citenamefont
  {Burrage}, \citenamefont {Coleman}, \citenamefont {Elertas}, \citenamefont
  {Ellis}, \citenamefont {Foot}, \citenamefont {Gibson}, \citenamefont
  {Haehnelt}, \citenamefont {Harte}, \citenamefont {Hedges}, \citenamefont
  {Hobson}, \citenamefont {Holynski}, \citenamefont {Jones}, \citenamefont
  {Langlois}, \citenamefont {Lellouch}, \citenamefont {Lewicki}, \citenamefont
  {Maiolino}, \citenamefont {Majewski}, \citenamefont {Malik}, \citenamefont
  {{March-Russell}}, \citenamefont {McCabe}, \citenamefont {Newbold},
  \citenamefont {Sauer}, \citenamefont {Schneider}, \citenamefont {Shipsey},
  \citenamefont {Singh}, \citenamefont {Uchida}, \citenamefont {Valenzuela},
  \citenamefont {van~der Grinten}, \citenamefont {Vaskonen}, \citenamefont
  {Vossebeld}, \citenamefont {Weatherill},\ and\ \citenamefont
  {Wilmut}}]{badurina_aion_2020}%
  \BibitemOpen
  \bibfield  {author} {\bibinfo {author} {\bibfnamefont {L.}~\bibnamefont
  {Badurina}}, \bibinfo {author} {\bibfnamefont {E.}~\bibnamefont {Bentine}},
  \bibinfo {author} {\bibfnamefont {D.}~\bibnamefont {Blas}}, \bibinfo {author}
  {\bibfnamefont {K.}~\bibnamefont {Bongs}}, \bibinfo {author} {\bibfnamefont
  {D.}~\bibnamefont {Bortoletto}}, \bibinfo {author} {\bibfnamefont
  {T.}~\bibnamefont {Bowcock}}, \bibinfo {author} {\bibfnamefont
  {K.}~\bibnamefont {Bridges}}, \bibinfo {author} {\bibfnamefont
  {W.}~\bibnamefont {Bowden}}, \bibinfo {author} {\bibfnamefont
  {O.}~\bibnamefont {Buchmueller}}, \bibinfo {author} {\bibfnamefont
  {C.}~\bibnamefont {Burrage}} et al.,\ }
  \bibfield {title} {\bibinfo {title} \textit{AION: an atom interferometer observatory and network},\ }\href
  {https://doi.org/10.1088/1475-7516/2020/05/011} {\bibfield  {journal}
  {\bibinfo  {journal} {J. Cosmol. Astropart. Phys.}\ }\textbf {\bibinfo
  {volume} {2020}}\bibinfo  {number} { (05)},\ \bibinfo {pages}
  {011}}\BibitemShut {NoStop}%
\bibitem [{\citenamefont {Canuel}\ \emph {et~al.}(2020)\citenamefont {Canuel},
  \citenamefont {Abend}, \citenamefont {{Amaro-Seoane}}, \citenamefont
  {Badaracco}, \citenamefont {Beaufils}, \citenamefont {Bertoldi},
  \citenamefont {Bongs}, \citenamefont {Bouyer}, \citenamefont {Braxmaier},
  \citenamefont {Chaibi}, \citenamefont {Christensen}, \citenamefont {Fitzek},
  \citenamefont {Flouris}, \citenamefont {Gaaloul}, \citenamefont {Gaffet},
  \citenamefont {Alzar}, \citenamefont {Geiger}, \citenamefont
  {{Guellati-Khelifa}}, \citenamefont {Hammerer}, \citenamefont {Harms},
  \citenamefont {Hinderer}, \citenamefont {Holynski}, \citenamefont {Junca},
  \citenamefont {Katsanevas}, \citenamefont {Klempt}, \citenamefont
  {Kozanitis}, \citenamefont {Krutzik}, \citenamefont {Landragin},
  \citenamefont {Roche}, \citenamefont {Leykauf}, \citenamefont {Lien},
  \citenamefont {Loriani}, \citenamefont {Merlet}, \citenamefont {Merzougui},
  \citenamefont {Nofrarias}, \citenamefont {Papadakos}, \citenamefont {dos
  Santos}, \citenamefont {Peters}, \citenamefont {Plexousakis}, \citenamefont
  {Prevedelli}, \citenamefont {Rasel}, \citenamefont {Rogister}, \citenamefont
  {Rosat}, \citenamefont {Roura}, \citenamefont {Sabulsky}, \citenamefont
  {Schkolnik}, \citenamefont {Schlippert}, \citenamefont {Schubert},
  \citenamefont {Sidorenkov}, \citenamefont {Siem{\ss}}, \citenamefont
  {Sopuerta}, \citenamefont {Sorrentino}, \citenamefont {Struckmann},
  \citenamefont {Tino}, \citenamefont {Tsagkatakis}, \citenamefont
  {Vicer{\'e}}, \citenamefont {von Klitzing}, \citenamefont {Woerner},\ and\
  \citenamefont {Zou}}]{canuel_elgar_2020}%
  \BibitemOpen
\bibfield  {number} {  }\bibfield  {author} {\bibinfo {author} {\bibfnamefont
  {B.}~\bibnamefont {Canuel}} et al.,\ }
  \bibfield {title} {\bibinfo {title} \textit{ELGAR — a European Laboratory for Gravitation
and Atom-interferometric Research},\ }\href
  {https://doi.org/10.1088/1361-6382/aba80e} {\bibfield  {journal} {\bibinfo
  {journal} {Class. Quantum Grav.}\ }\textbf {\bibinfo {volume} {37}},\
  \bibinfo {pages} {225017} (\bibinfo {year} {2020})}\BibitemShut {NoStop}%
\bibitem [{\citenamefont {Tino}\ \emph {et~al.}(2019)\citenamefont {Tino},
  \citenamefont {Bassi}, \citenamefont {Bianco}, \citenamefont {Bongs},
  \citenamefont {Bouyer}, \citenamefont {Cacciapuoti}, \citenamefont
  {Capozziello}, \citenamefont {Chen}, \citenamefont {Chiofalo}, \citenamefont
  {Derevianko}, \citenamefont {Ertmer}, \citenamefont {Gaaloul}, \citenamefont
  {Gill}, \citenamefont {Graham}, \citenamefont {Hogan}, \citenamefont {Iess},
  \citenamefont {Kasevich}, \citenamefont {Katori}, \citenamefont {Klempt},
  \citenamefont {Lu}, \citenamefont {Ma}, \citenamefont {M{\"u}ller},
  \citenamefont {Newbury}, \citenamefont {Oates}, \citenamefont {Peters},
  \citenamefont {Poli}, \citenamefont {Rasel}, \citenamefont {Rosi},
  \citenamefont {Roura}, \citenamefont {Salomon}, \citenamefont {Schiller},
  \citenamefont {Schleich}, \citenamefont {Schlippert}, \citenamefont
  {Schreck}, \citenamefont {Schubert}, \citenamefont {Sorrentino},
  \citenamefont {Sterr}, \citenamefont {Thomsen}, \citenamefont {Vallone},
  \citenamefont {Vetrano}, \citenamefont {Villoresi}, \citenamefont {{von
  Klitzing}}, \citenamefont {Wilkowski}, \citenamefont {Wolf}, \citenamefont
  {Ye}, \citenamefont {Yu},\ and\ \citenamefont {Zhan}}]{tino_sage_2019}%
  \BibitemOpen
  \bibfield  {author} {\bibinfo {author} {\bibfnamefont {G.~M.}\ \bibnamefont
  {Tino}} et al.,\
  }
  \bibfield {title} {\bibinfo {title} \textit{SAGE: A proposal for a space atomic gravity explorer},\ }\href {https://doi.org/10.1140/epjd/e2019-100324-6} {\bibfield  {journal}
  {\bibinfo  {journal} {Eur. Phys. J. D}\ }\textbf {\bibinfo {volume} {73}},\
  \bibinfo {pages} {228} (\bibinfo {year} {2019})}\BibitemShut {NoStop}%
\bibitem [{\citenamefont {Canuel}\ \emph {et~al.}(2018)\citenamefont {Canuel},
  \citenamefont {Bertoldi}, \citenamefont {Amand}, \citenamefont {{Pozzo di
  Borgo}}, \citenamefont {Chantrait}, \citenamefont {Danquigny}, \citenamefont
  {Dovale~{\'A}lvarez}, \citenamefont {Fang}, \citenamefont {Freise},
  \citenamefont {Geiger}, \citenamefont {Gillot}, \citenamefont {Henry},
  \citenamefont {Hinderer}, \citenamefont {Holleville}, \citenamefont {Junca},
  \citenamefont {Lef{\`e}vre}, \citenamefont {Merzougui}, \citenamefont
  {Mielec}, \citenamefont {Monfret}, \citenamefont {Pelisson}, \citenamefont
  {Prevedelli}, \citenamefont {Reynaud}, \citenamefont {Riou}, \citenamefont
  {Rogister}, \citenamefont {Rosat}, \citenamefont {Cormier}, \citenamefont
  {Landragin}, \citenamefont {Chaibi}, \citenamefont {Gaffet},\ and\
  \citenamefont {Bouyer}}]{canuel_exploring_2018}%
  \BibitemOpen
  \bibfield  {author} {\bibinfo {author} {\bibfnamefont {B.}~\bibnamefont
  {Canuel}} et al.,\ }
  \bibfield {title} {\bibinfo {title} \textit{Exploring gravity with the MIGA large scale atom interferometer},\ }\href
  {https://doi.org/10.1038/s41598-018-32165-z} {\bibfield  {journal} {\bibinfo
  {journal} {Sci Rep}\ }\textbf {\bibinfo {volume} {8}},\ \bibinfo {pages}
  {14064} (\bibinfo {year} {2018})}\BibitemShut {NoStop}%
\bibitem [{\citenamefont {Hogan}\ and\ \citenamefont
  {Kasevich}(2016)}]{hogan_atominterferometric_2016}%
  \BibitemOpen
  \bibfield  {author} {\bibinfo {author} {\bibfnamefont {J.~M.}\ \bibnamefont
  {Hogan}}\ and\ \bibinfo {author} {\bibfnamefont {M.~A.}\ \bibnamefont
  {Kasevich}},\ }\bibfield {title} {\bibinfo {title} \textit{Atom-interferometric gravitational-wave detection using heterodyne laser links
},\ }\href {https://doi.org/10.1103/PhysRevA.94.033632} {\bibfield
  {journal} {\bibinfo  {journal} {Phys. Rev. A}\ }\textbf {\bibinfo {volume}
  {94}},\ \bibinfo {pages} {033632} (\bibinfo {year} {2016})}\BibitemShut
  {NoStop}%
\bibitem [{\citenamefont {Hogan}\ \emph {et~al.}(2011)\citenamefont {Hogan},
  \citenamefont {Johnson}, \citenamefont {Dickerson}, \citenamefont {Kovachy},
  \citenamefont {Sugarbaker}, \citenamefont {Chiow}, \citenamefont {Graham},
  \citenamefont {Kasevich}, \citenamefont {Saif}, \citenamefont {Rajendran},
  \citenamefont {Bouyer}, \citenamefont {Seery}, \citenamefont {Feinberg},\
  and\ \citenamefont {{Keski-Kuha}}}]{hogan_atomic_2011}%
  \BibitemOpen
  \bibfield  {author} {\bibinfo {author} {\bibfnamefont {J.~M.}\ \bibnamefont
  {Hogan}}, \bibinfo {author} {\bibfnamefont {D.~M.~S.}\ \bibnamefont
  {Johnson}}, \bibinfo {author} {\bibfnamefont {S.}~\bibnamefont {Dickerson}},
  \bibinfo {author} {\bibfnamefont {T.}~\bibnamefont {Kovachy}}, \bibinfo
  {author} {\bibfnamefont {A.}~\bibnamefont {Sugarbaker}}, \bibinfo {author}
  {\bibfnamefont {S.-w.}\ \bibnamefont {Chiow}}, \bibinfo {author}
  {\bibfnamefont {P.~W.}\ \bibnamefont {Graham}}, \bibinfo {author}
  {\bibfnamefont {M.~A.}\ \bibnamefont {Kasevich}}, \bibinfo {author}
  {\bibfnamefont {B.}~\bibnamefont {Saif}}, \bibinfo {author} {\bibfnamefont
  {S.}~\bibnamefont {Rajendran}} et al.,\ }
  \bibfield {title} {\bibinfo {title} \textit{An atomic gravitational wave interferometric sensor in low earth orbit (AGIS-LEO)
},\ }\href {https://doi.org/10.1007/s10714-011-1182-x}
  {\bibfield  {journal} {\bibinfo  {journal} {Gen Relativ Gravit}\ }\textbf
  {\bibinfo {volume} {43}},\ \bibinfo {pages} {1953} (\bibinfo {year}
  {2011})}\BibitemShut {NoStop}%
\bibitem [{\citenamefont {Dimopoulos}\ \emph
  {et~al.}(2008{\natexlab{b}})\citenamefont {Dimopoulos}, \citenamefont
  {Graham}, \citenamefont {Hogan}, \citenamefont {Kasevich},\ and\
  \citenamefont {Rajendran}}]{dimopoulos_atomic_2008}%
  \BibitemOpen
  \bibfield  {author} {\bibinfo {author} {\bibfnamefont {S.}~\bibnamefont
  {Dimopoulos}}, \bibinfo {author} {\bibfnamefont {P.~W.}\ \bibnamefont
  {Graham}}, \bibinfo {author} {\bibfnamefont {J.~M.}\ \bibnamefont {Hogan}},
  \bibinfo {author} {\bibfnamefont {M.~A.}\ \bibnamefont {Kasevich}},\ and\
  \bibinfo {author} {\bibfnamefont {S.}~\bibnamefont {Rajendran}},\ }
  \bibfield {title} {\bibinfo {title} \textit{Atomic gravitational wave interferometric sensor
},\ }\href
  {https://doi.org/10.1103/PhysRevD.78.122002} {\bibfield  {journal} {\bibinfo
  {journal} {Phys. Rev. D}\ }\textbf {\bibinfo {volume} {78}},\ \bibinfo
  {pages} {122002} (\bibinfo {year} {2008}{\natexlab{b}})}\BibitemShut
  {NoStop}%
\bibitem [{\citenamefont {Feldmann}\ \emph {et~al.}(2023)\citenamefont
  {Feldmann}, \citenamefont {Anders}, \citenamefont {Idel}, \citenamefont
  {Schubert}, \citenamefont {Schlippert}, \citenamefont {Santos}, \citenamefont
  {Rasel},\ and\ \citenamefont {Klempt}}]{feldmann_optimal_2023}%
  \BibitemOpen
  \bibfield  {author} {\bibinfo {author} {\bibfnamefont {P.}~\bibnamefont
  {Feldmann}}, \bibinfo {author} {\bibfnamefont {F.}~\bibnamefont {Anders}},
  \bibinfo {author} {\bibfnamefont {A.}~\bibnamefont {Idel}}, \bibinfo {author}
  {\bibfnamefont {C.}~\bibnamefont {Schubert}}, \bibinfo {author}
  {\bibfnamefont {D.}~\bibnamefont {Schlippert}}, \bibinfo {author}
  {\bibfnamefont {L.}~\bibnamefont {Santos}}, \bibinfo {author} {\bibfnamefont
  {E.~M.}\ \bibnamefont {Rasel}},\ and\ \bibinfo {author} {\bibfnamefont
  {C.}~\bibnamefont {Klempt}},\ }
  \bibfield {title} {\bibinfo {title} \textit{Optimal squeezing for high-precision atom interferometers
},\ }
  \href
  {https://doi.org/10.48550/arXiv.2311.10241} {\bibfield  {journal} {\bibinfo
  {journal} {arXiv}\ ,\ \bibinfo {pages} {2311.10241}} (\bibinfo {year}
  {2023})}\BibitemShut {NoStop}%
\bibitem [{\citenamefont {Pezz{\`e}}\ \emph {et~al.}(2018)\citenamefont
  {Pezz{\`e}}, \citenamefont {Smerzi}, \citenamefont {Oberthaler},
  \citenamefont {Schmied},\ and\ \citenamefont
  {Treutlein}}]{pezze_quantum_2018}%
  \BibitemOpen
  \bibfield  {author} {\bibinfo {author} {\bibfnamefont {L.}~\bibnamefont
  {Pezz{\`e}}}, \bibinfo {author} {\bibfnamefont {A.}~\bibnamefont {Smerzi}},
  \bibinfo {author} {\bibfnamefont {M.~K.}\ \bibnamefont {Oberthaler}},
  \bibinfo {author} {\bibfnamefont {R.}~\bibnamefont {Schmied}},\ and\ \bibinfo
  {author} {\bibfnamefont {P.}~\bibnamefont {Treutlein}},\ }
  \bibfield {title} {\bibinfo {title} \textit{Quantum metrology with nonclassical states of atomic ensembles
},\ }
\href
  {https://doi.org/10.1103/RevModPhys.90.035005} {\bibfield  {journal}
  {\bibinfo  {journal} {Rev. Mod. Phys.}\ }\textbf {\bibinfo {volume} {90}},\
  \bibinfo {pages} {035005} (\bibinfo {year} {2018})}\BibitemShut {NoStop}%
\bibitem [{\citenamefont {Bonneau}\ \emph {et~al.}(2013)\citenamefont
  {Bonneau}, \citenamefont {Ruaudel}, \citenamefont {Lopes}, \citenamefont
  {Jaskula}, \citenamefont {Aspect}, \citenamefont {Boiron},\ and\
  \citenamefont {Westbrook}}]{bonneau_tunable_2013}%
  \BibitemOpen
  \bibfield  {author} {\bibinfo {author} {\bibfnamefont {M.}~\bibnamefont
  {Bonneau}}, \bibinfo {author} {\bibfnamefont {J.}~\bibnamefont {Ruaudel}},
  \bibinfo {author} {\bibfnamefont {R.}~\bibnamefont {Lopes}}, \bibinfo
  {author} {\bibfnamefont {J.-C.}\ \bibnamefont {Jaskula}}, \bibinfo {author}
  {\bibfnamefont {A.}~\bibnamefont {Aspect}}, \bibinfo {author} {\bibfnamefont
  {D.}~\bibnamefont {Boiron}},\ and\ \bibinfo {author} {\bibfnamefont {C.~I.}\
  \bibnamefont {Westbrook}},\ }
  \bibfield {title} {\bibinfo {title} \textit{Tunable source of correlated atom beams
},\ }
\href
  {https://doi.org/10.1103/PhysRevA.87.061603} {\bibfield  {journal} {\bibinfo
  {journal} {Phys. Rev. A}\ }\textbf {\bibinfo {volume} {87}},\ \bibinfo
  {pages} {061603} (\bibinfo {year} {2013})}\BibitemShut {NoStop}%
\bibitem [{\citenamefont {Dussarrat}\ \emph {et~al.}(2017)\citenamefont
  {Dussarrat}, \citenamefont {Perrier}, \citenamefont {Imanaliev},
  \citenamefont {Lopes}, \citenamefont {Aspect}, \citenamefont {Cheneau},
  \citenamefont {Boiron},\ and\ \citenamefont
  {Westbrook}}]{dussarrat_twoparticle_2017}%
  \BibitemOpen
  \bibfield  {author} {\bibinfo {author} {\bibfnamefont {P.}~\bibnamefont
  {Dussarrat}}, \bibinfo {author} {\bibfnamefont {M.}~\bibnamefont {Perrier}},
  \bibinfo {author} {\bibfnamefont {A.}~\bibnamefont {Imanaliev}}, \bibinfo
  {author} {\bibfnamefont {R.}~\bibnamefont {Lopes}}, \bibinfo {author}
  {\bibfnamefont {A.}~\bibnamefont {Aspect}}, \bibinfo {author} {\bibfnamefont
  {M.}~\bibnamefont {Cheneau}}, \bibinfo {author} {\bibfnamefont
  {D.}~\bibnamefont {Boiron}},\ and\ \bibinfo {author} {\bibfnamefont {C.~I.}\
  \bibnamefont {Westbrook}},\ }
  \bibfield {title} {\bibinfo {title} \textit{Two-Particle Four-Mode Interferometer for Atoms
},\ }
\href
  {https://doi.org/10.1103/PhysRevLett.119.173202} {\bibfield  {journal}
  {\bibinfo  {journal} {Phys. Rev. Lett.}\ }\textbf {\bibinfo {volume} {119}},\
  \bibinfo {pages} {173202} (\bibinfo {year} {2017})}\BibitemShut {NoStop}%
\bibitem [{\citenamefont {B{\"u}cker}\ \emph {et~al.}(2011)\citenamefont
  {B{\"u}cker}, \citenamefont {Grond}, \citenamefont {Manz}, \citenamefont
  {Berrada}, \citenamefont {Betz}, \citenamefont {Koller}, \citenamefont
  {Hohenester}, \citenamefont {Schumm}, \citenamefont {Perrin},\ and\
  \citenamefont {Schmiedmayer}}]{bucker_Twinatom_2011}%
  \BibitemOpen
  \bibfield  {author} {\bibinfo {author} {\bibfnamefont {R.}~\bibnamefont
  {B{\"u}cker}}, \bibinfo {author} {\bibfnamefont {J.}~\bibnamefont {Grond}},
  \bibinfo {author} {\bibfnamefont {S.}~\bibnamefont {Manz}}, \bibinfo {author}
  {\bibfnamefont {T.}~\bibnamefont {Berrada}}, \bibinfo {author} {\bibfnamefont
  {T.}~\bibnamefont {Betz}}, \bibinfo {author} {\bibfnamefont {C.}~\bibnamefont
  {Koller}}, \bibinfo {author} {\bibfnamefont {U.}~\bibnamefont {Hohenester}},
  \bibinfo {author} {\bibfnamefont {T.}~\bibnamefont {Schumm}}, \bibinfo
  {author} {\bibfnamefont {A.}~\bibnamefont {Perrin}},\ and\ \bibinfo {author}
  {\bibfnamefont {J.}~\bibnamefont {Schmiedmayer}},\ }
  \bibfield {title} {\bibinfo {title} \textit{Twin-atom beams
},\ }
\href
  {https://doi.org/10.1038/nphys1992} {\bibfield  {journal} {\bibinfo
  {journal} {Nature Phys}\ }\textbf {\bibinfo {volume} {7}},\ \bibinfo {pages}
  {608} (\bibinfo {year} {2011})}\BibitemShut {NoStop}%
\bibitem [{\citenamefont {Anders}\ \emph {et~al.}(2021)\citenamefont {Anders},
  \citenamefont {Idel}, \citenamefont {Feldmann}, \citenamefont {Bondarenko},
  \citenamefont {Loriani}, \citenamefont {Lange}, \citenamefont {Peise},
  \citenamefont {Gersemann}, \citenamefont {{Meyer-Hoppe}}, \citenamefont
  {Abend}, \citenamefont {Gaaloul}, \citenamefont {Schubert}, \citenamefont
  {Schlippert}, \citenamefont {Santos}, \citenamefont {Rasel},\ and\
  \citenamefont {Klempt}}]{anders_momentum_2021}%
  \BibitemOpen
  \bibfield  {author} {\bibinfo {author} {\bibfnamefont {F.}~\bibnamefont
  {Anders}}, \bibinfo {author} {\bibfnamefont {A.}~\bibnamefont {Idel}},
  \bibinfo {author} {\bibfnamefont {P.}~\bibnamefont {Feldmann}}, \bibinfo
  {author} {\bibfnamefont {D.}~\bibnamefont {Bondarenko}}, \bibinfo {author}
  {\bibfnamefont {S.}~\bibnamefont {Loriani}}, \bibinfo {author} {\bibfnamefont
  {K.}~\bibnamefont {Lange}}, \bibinfo {author} {\bibfnamefont
  {J.}~\bibnamefont {Peise}}, \bibinfo {author} {\bibfnamefont
  {M.}~\bibnamefont {Gersemann}}, \bibinfo {author} {\bibfnamefont
  {B.}~\bibnamefont {{Meyer-Hoppe}}}, \bibinfo {author} {\bibfnamefont
  {S.}~\bibnamefont {Abend}} et al.,\ }
  \bibfield {title} {\bibinfo {title} \textit{Momentum Entanglement for Atom Interferometry
},\ }
\href
  {https://doi.org/10.1103/PhysRevLett.127.140402} {\bibfield  {journal}
  {\bibinfo  {journal} {Phys. Rev. Lett.}\ }\textbf {\bibinfo {volume} {127}},\
  \bibinfo {pages} {140402} (\bibinfo {year} {2021})}\BibitemShut {NoStop}%
\bibitem [{\citenamefont {Malia}\ \emph {et~al.}(2022)\citenamefont {Malia},
  \citenamefont {Wu}, \citenamefont {{Mart{\'i}nez-Rinc{\'o}n}},\ and\
  \citenamefont {Kasevich}}]{malia_distributed_2022}%
  \BibitemOpen
  \bibfield  {author} {\bibinfo {author} {\bibfnamefont {B.~K.}\ \bibnamefont
  {Malia}}, \bibinfo {author} {\bibfnamefont {Y.}~\bibnamefont {Wu}}, \bibinfo
  {author} {\bibfnamefont {J.}~\bibnamefont {{Mart{\'i}nez-Rinc{\'o}n}}},\ and\
  \bibinfo {author} {\bibfnamefont {M.~A.}\ \bibnamefont {Kasevich}},\ }
  \bibfield {title} {\bibinfo {title} \textit{Distributed quantum sensing with mode-entangled spin-squeezed atomic states
},\ }
\href
  {https://doi.org/10.1038/s41586-022-05363-z} {\bibfield  {journal} {\bibinfo
  {journal} {Nature}\ }\textbf {\bibinfo {volume} {612}},\ \bibinfo {pages}
  {661} (\bibinfo {year} {2022})}\BibitemShut {NoStop}%
\bibitem [{\citenamefont {Greve}\ \emph {et~al.}(2022)\citenamefont {Greve},
  \citenamefont {Luo}, \citenamefont {Wu},\ and\ \citenamefont
  {Thompson}}]{greve_entanglementenhanced_2022}%
  \BibitemOpen
  \bibfield  {author} {\bibinfo {author} {\bibfnamefont {G.~P.}\ \bibnamefont
  {Greve}}, \bibinfo {author} {\bibfnamefont {C.}~\bibnamefont {Luo}}, \bibinfo
  {author} {\bibfnamefont {B.}~\bibnamefont {Wu}},\ and\ \bibinfo {author}
  {\bibfnamefont {J.~K.}\ \bibnamefont {Thompson}},\ }
  \bibfield {title} {\bibinfo {title} \textit{Entanglement-enhanced matter-wave interferometry in a high-finesse cavity
},\ }
\href
  {https://doi.org/10.1038/s41586-022-05197-9} {\bibfield  {journal} {\bibinfo
  {journal} {Nature}\ }\textbf {\bibinfo {volume} {610}},\ \bibinfo {pages}
  {472} (\bibinfo {year} {2022})}\BibitemShut {NoStop}%
\bibitem [{\citenamefont {Deppner}\ \emph {et~al.}(2021)\citenamefont
  {Deppner}, \citenamefont {Herr}, \citenamefont {Cornelius}, \citenamefont
  {Stromberger}, \citenamefont {Sternke}, \citenamefont {Grzeschik},
  \citenamefont {Grote}, \citenamefont {Rudolph}, \citenamefont {Herrmann},
  \citenamefont {Krutzik}, \citenamefont {Wenzlawski}, \citenamefont {Corgier},
  \citenamefont {Charron}, \citenamefont {{Gu{\'e}ry-Odelin}}, \citenamefont
  {Gaaloul}, \citenamefont {L{\"a}mmerzahl}, \citenamefont {Peters},
  \citenamefont {Windpassinger},\ and\ \citenamefont
  {Rasel}}]{deppner_CollectiveMode_2021}%
  \BibitemOpen
  \bibfield  {author} {\bibinfo {author} {\bibfnamefont {C.}~\bibnamefont
  {Deppner}}, \bibinfo {author} {\bibfnamefont {W.}~\bibnamefont {Herr}},
  \bibinfo {author} {\bibfnamefont {M.}~\bibnamefont {Cornelius}}, \bibinfo
  {author} {\bibfnamefont {P.}~\bibnamefont {Stromberger}}, \bibinfo {author}
  {\bibfnamefont {T.}~\bibnamefont {Sternke}}, \bibinfo {author} {\bibfnamefont
  {C.}~\bibnamefont {Grzeschik}}, \bibinfo {author} {\bibfnamefont
  {A.}~\bibnamefont {Grote}}, \bibinfo {author} {\bibfnamefont
  {J.}~\bibnamefont {Rudolph}}, \bibinfo {author} {\bibfnamefont
  {S.}~\bibnamefont {Herrmann}}, \bibinfo {author} {\bibfnamefont
  {M.}~\bibnamefont {Krutzik}} et al.,\ }
  \bibfield {title} {\bibinfo {title} \textit{Collective-Mode Enhanced Matter-Wave Optics
},\ }
\href
  {https://doi.org/10.1103/PhysRevLett.127.100401} {\bibfield  {journal}
  {\bibinfo  {journal} {Phys. Rev. Lett.}\ }\textbf {\bibinfo {volume} {127}},\
  \bibinfo {pages} {100401} (\bibinfo {year} {2021})}\BibitemShut {NoStop}%
\bibitem [{\citenamefont {Chiow}\ \emph {et~al.}(2011)\citenamefont {Chiow},
  \citenamefont {Kovachy}, \citenamefont {Chien},\ and\ \citenamefont
  {Kasevich}}]{chiow_102hhbark_2011}%
  \BibitemOpen
  \bibfield  {author} {\bibinfo {author} {\bibfnamefont {S.-w.}\ \bibnamefont
  {Chiow}}, \bibinfo {author} {\bibfnamefont {T.}~\bibnamefont {Kovachy}},
  \bibinfo {author} {\bibfnamefont {H.-C.}\ \bibnamefont {Chien}},\ and\
  \bibinfo {author} {\bibfnamefont {M.~A.}\ \bibnamefont {Kasevich}},\ }
  \bibfield {title} {\bibinfo {title} \textit{$102\hbar$k Large Area Atom Interferometers
},\ }
\href
  {https://doi.org/10.1103/PhysRevLett.107.130403} {\bibfield  {journal}
  {\bibinfo  {journal} {Phys. Rev. Lett.}\ }\textbf {\bibinfo {volume} {107}},\
  \bibinfo {pages} {130403} (\bibinfo {year} {2011})}\BibitemShut {NoStop}%
\bibitem [{\citenamefont {Debs}\ \emph {et~al.}(2011)\citenamefont {Debs},
  \citenamefont {Altin}, \citenamefont {Barter}, \citenamefont {D{\"o}ring},
  \citenamefont {Dennis}, \citenamefont {McDonald}, \citenamefont {Anderson},
  \citenamefont {Close},\ and\ \citenamefont {Robins}}]{debs_Coldatom_2011}%
  \BibitemOpen
  \bibfield  {author} {\bibinfo {author} {\bibfnamefont {J.~E.}\ \bibnamefont
  {Debs}}, \bibinfo {author} {\bibfnamefont {P.~A.}\ \bibnamefont {Altin}},
  \bibinfo {author} {\bibfnamefont {T.~H.}\ \bibnamefont {Barter}}, \bibinfo
  {author} {\bibfnamefont {D.}~\bibnamefont {D{\"o}ring}}, \bibinfo {author}
  {\bibfnamefont {G.~R.}\ \bibnamefont {Dennis}}, \bibinfo {author}
  {\bibfnamefont {G.}~\bibnamefont {McDonald}}, \bibinfo {author}
  {\bibfnamefont {R.~P.}\ \bibnamefont {Anderson}}, \bibinfo {author}
  {\bibfnamefont {J.~D.}\ \bibnamefont {Close}},\ and\ \bibinfo {author}
  {\bibfnamefont {N.~P.}\ \bibnamefont {Robins}},\ }
  \bibfield {title} {\bibinfo {title} \textit{Cold-atom gravimetry with a Bose-Einstein condensate
},\ }
\href
  {https://doi.org/10.1103/PhysRevA.84.033610} {\bibfield  {journal} {\bibinfo
  {journal} {Phys. Rev. A}\ }\textbf {\bibinfo {volume} {84}},\ \bibinfo
  {pages} {033610} (\bibinfo {year} {2011})}\BibitemShut {NoStop}%
\bibitem [{\citenamefont {Gebbe}\ \emph {et~al.}(2021)\citenamefont {Gebbe},
  \citenamefont {Siem{\ss}}, \citenamefont {Gersemann}, \citenamefont
  {M{\"u}ntinga}, \citenamefont {Herrmann}, \citenamefont {L{\"a}mmerzahl},
  \citenamefont {Ahlers}, \citenamefont {Gaaloul}, \citenamefont {Schubert},
  \citenamefont {Hammerer}, \citenamefont {Abend},\ and\ \citenamefont
  {Rasel}}]{gebbe_Twinlattice_2021}%
  \BibitemOpen
  \bibfield  {author} {\bibinfo {author} {\bibfnamefont {M.}~\bibnamefont
  {Gebbe}}, \bibinfo {author} {\bibfnamefont {J.-N.}\ \bibnamefont
  {Siem{\ss}}}, \bibinfo {author} {\bibfnamefont {M.}~\bibnamefont
  {Gersemann}}, \bibinfo {author} {\bibfnamefont {H.}~\bibnamefont
  {M{\"u}ntinga}}, \bibinfo {author} {\bibfnamefont {S.}~\bibnamefont
  {Herrmann}}, \bibinfo {author} {\bibfnamefont {C.}~\bibnamefont
  {L{\"a}mmerzahl}}, \bibinfo {author} {\bibfnamefont {H.}~\bibnamefont
  {Ahlers}}, \bibinfo {author} {\bibfnamefont {N.}~\bibnamefont {Gaaloul}},
  \bibinfo {author} {\bibfnamefont {C.}~\bibnamefont {Schubert}}, \bibinfo
  {author} {\bibfnamefont {K.}~\bibnamefont {Hammerer}}, \bibinfo {author}
  {\bibfnamefont {S.}~\bibnamefont {Abend}},\ and\ \bibinfo {author}
  {\bibfnamefont {E.~M.}\ \bibnamefont {Rasel}},\ }
  \bibfield {title} {\bibinfo {title} \textit{Twin-lattice atom interferometry
},\ }
\href
  {https://doi.org/10.1038/s41467-021-22823-8} {\bibfield  {journal} {\bibinfo
  {journal} {Nat Commun}\ }\textbf {\bibinfo {volume} {12}},\ \bibinfo {pages}
  {2544} (\bibinfo {year} {2021})}\BibitemShut {NoStop}%
\bibitem [{\citenamefont {Szigeti}\ \emph {et~al.}(2020)\citenamefont
  {Szigeti}, \citenamefont {Nolan}, \citenamefont {Close},\ and\ \citenamefont
  {Haine}}]{szigetti_high_precision}%
  \BibitemOpen
  \bibfield  {author} {\bibinfo {author} {\bibfnamefont {S.~S.}\ \bibnamefont
  {Szigeti}}, \bibinfo {author} {\bibfnamefont {S.~P.}\ \bibnamefont {Nolan}},
  \bibinfo {author} {\bibfnamefont {J.~D.}\ \bibnamefont {Close}},\ and\
  \bibinfo {author} {\bibfnamefont {S.~A.}\ \bibnamefont {Haine}},\ }
  \bibfield {title} {\bibinfo {title} \textit{High-Precision Quantum-Enhanced Gravimetry with a Bose-Einstein Condensate
},\ }
\href
  {https://doi.org/10.1103/PhysRevLett.125.100402} {\bibfield  {journal}
  {\bibinfo  {journal} {Phys. Rev. Lett.}\ }\textbf {\bibinfo {volume} {125}},\
  \bibinfo {pages} {100402} (\bibinfo {year} {2020})}\BibitemShut {NoStop}%
\bibitem [{\citenamefont {Corgier}\ \emph {et~al.}(2021)\citenamefont
  {Corgier}, \citenamefont {Gaaloul}, \citenamefont {Smerzi},\ and\
  \citenamefont {Pezz\`e}}]{corgier_delta_kick}%
  \BibitemOpen
  \bibfield  {author} {\bibinfo {author} {\bibfnamefont {R.}~\bibnamefont
  {Corgier}}, \bibinfo {author} {\bibfnamefont {N.}~\bibnamefont {Gaaloul}},
  \bibinfo {author} {\bibfnamefont {A.}~\bibnamefont {Smerzi}},\ and\ \bibinfo
  {author} {\bibfnamefont {L.}~\bibnamefont {Pezz\`e}},\ }
  \bibfield {title} {\bibinfo {title} \textit{Delta-Kick Squeezing
},\ }
\href
  {https://doi.org/10.1103/PhysRevLett.127.183401} {\bibfield  {journal}
  {\bibinfo  {journal} {Phys. Rev. Lett.}\ }\textbf {\bibinfo {volume} {127}},\
  \bibinfo {pages} {183401} (\bibinfo {year} {2021})}\BibitemShut {NoStop}%
\bibitem [{\citenamefont {Hamley}\ \emph {et~al.}(2012)\citenamefont {Hamley},
  \citenamefont {Gerving}, \citenamefont {Hoang}, \citenamefont {Bookjans},\
  and\ \citenamefont {Chapman}}]{hamley_spinnematic_2012}%
  \BibitemOpen
  \bibfield  {author} {\bibinfo {author} {\bibfnamefont {C.~D.}\ \bibnamefont
  {Hamley}}, \bibinfo {author} {\bibfnamefont {C.~S.}\ \bibnamefont {Gerving}},
  \bibinfo {author} {\bibfnamefont {T.~M.}\ \bibnamefont {Hoang}}, \bibinfo
  {author} {\bibfnamefont {E.~M.}\ \bibnamefont {Bookjans}},\ and\ \bibinfo
  {author} {\bibfnamefont {M.~S.}\ \bibnamefont {Chapman}},\ }
  \bibfield {title} {\bibinfo {title} \textit{Spin-nematic squeezed vacuum in a quantum gas
},\ }
\href
  {https://doi.org/10.1038/nphys2245} {\bibfield  {journal} {\bibinfo
  {journal} {Nature Phys}\ }\textbf {\bibinfo {volume} {8}},\ \bibinfo {pages}
  {305} (\bibinfo {year} {2012})}\BibitemShut {NoStop}%
\bibitem [{\citenamefont {Peise}\ \emph {et~al.}(2015)\citenamefont {Peise},
  \citenamefont {Kruse}, \citenamefont {Lange}, \citenamefont {L{\"u}cke},
  \citenamefont {Pezz{\`e}}, \citenamefont {Arlt}, \citenamefont {Ertmer},
  \citenamefont {Hammerer}, \citenamefont {Santos}, \citenamefont {Smerzi},\
  and\ \citenamefont {Klempt}}]{peise_satisfying_2015}%
  \BibitemOpen
  \bibfield  {author} {\bibinfo {author} {\bibfnamefont {J.}~\bibnamefont
  {Peise}}, \bibinfo {author} {\bibfnamefont {I.}~\bibnamefont {Kruse}},
  \bibinfo {author} {\bibfnamefont {K.}~\bibnamefont {Lange}}, \bibinfo
  {author} {\bibfnamefont {B.}~\bibnamefont {L{\"u}cke}}, \bibinfo {author}
  {\bibfnamefont {L.}~\bibnamefont {Pezz{\`e}}}, \bibinfo {author}
  {\bibfnamefont {J.}~\bibnamefont {Arlt}}, \bibinfo {author} {\bibfnamefont
  {W.}~\bibnamefont {Ertmer}}, \bibinfo {author} {\bibfnamefont
  {K.}~\bibnamefont {Hammerer}}, \bibinfo {author} {\bibfnamefont
  {L.}~\bibnamefont {Santos}}, \bibinfo {author} {\bibfnamefont
  {A.}~\bibnamefont {Smerzi}},\ and\ \bibinfo {author} {\bibfnamefont
  {C.}~\bibnamefont {Klempt}},\ }
  \bibfield {title} {\bibinfo {title} \textit{Satisfying the Einstein–Podolsky–Rosen criterion with massive particles
},\ }
\href {https://doi.org/10.1038/ncomms9984}
  {\bibfield  {journal} {\bibinfo  {journal} {Nat Commun}\ }\textbf {\bibinfo
  {volume} {6}},\ \bibinfo {pages} {8984} (\bibinfo {year} {2015})}\BibitemShut
  {NoStop}%
\bibitem [{\citenamefont {Kruse}\ \emph {et~al.}(2016)\citenamefont {Kruse},
  \citenamefont {Lange}, \citenamefont {Peise}, \citenamefont {L{\"u}cke},
  \citenamefont {Pezz{\`e}}, \citenamefont {Arlt}, \citenamefont {Ertmer},
  \citenamefont {Lisdat}, \citenamefont {Santos}, \citenamefont {Smerzi},\ and\
  \citenamefont {Klempt}}]{kruse_improvement_2016}%
  \BibitemOpen
  \bibfield  {author} {\bibinfo {author} {\bibfnamefont {I.}~\bibnamefont
  {Kruse}}, \bibinfo {author} {\bibfnamefont {K.}~\bibnamefont {Lange}},
  \bibinfo {author} {\bibfnamefont {J.}~\bibnamefont {Peise}}, \bibinfo
  {author} {\bibfnamefont {B.}~\bibnamefont {L{\"u}cke}}, \bibinfo {author}
  {\bibfnamefont {L.}~\bibnamefont {Pezz{\`e}}}, \bibinfo {author}
  {\bibfnamefont {J.}~\bibnamefont {Arlt}}, \bibinfo {author} {\bibfnamefont
  {W.}~\bibnamefont {Ertmer}}, \bibinfo {author} {\bibfnamefont
  {C.}~\bibnamefont {Lisdat}}, \bibinfo {author} {\bibfnamefont
  {L.}~\bibnamefont {Santos}}, \bibinfo {author} {\bibfnamefont
  {A.}~\bibnamefont {Smerzi}},\ and\ \bibinfo {author} {\bibfnamefont
  {C.}~\bibnamefont {Klempt}},\ }
  \bibfield {title} {\bibinfo {title} \textit{Improvement of an Atomic Clock using Squeezed Vacuum
},\ }
\href
  {https://doi.org/10.1103/PhysRevLett.117.143004} {\bibfield  {journal}
  {\bibinfo  {journal} {Phys. Rev. Lett.}\ }\textbf {\bibinfo {volume} {117}},\
  \bibinfo {pages} {143004} (\bibinfo {year} {2016})}\BibitemShut {NoStop}%
\bibitem [{\citenamefont {Ammann}\ and\ \citenamefont
  {Christensen}(1997)}]{ammann_delta_1997}%
  \BibitemOpen
  \bibfield  {author} {\bibinfo {author} {\bibfnamefont {H.}~\bibnamefont
  {Ammann}}\ and\ \bibinfo {author} {\bibfnamefont {N.}~\bibnamefont
  {Christensen}},\ }
  \bibfield {title} {\bibinfo {title} \textit{Delta Kick Cooling: A New Method for Cooling Atoms
},\ }
\href {https://doi.org/10.1103/PhysRevLett.78.2088}
  {\bibfield  {journal} {\bibinfo  {journal} {Phys. Rev. Lett.}\ }\textbf
  {\bibinfo {volume} {78}},\ \bibinfo {pages} {2088} (\bibinfo {year}
  {1997})}\BibitemShut {NoStop}%
\bibitem [{\citenamefont {Kunkel}\ \emph {et~al.}(2019)\citenamefont {Kunkel},
  \citenamefont {Pr{\"u}fer}, \citenamefont {Lannig}, \citenamefont
  {{Rosa-Medina}}, \citenamefont {Bonnin}, \citenamefont {G{\"a}rttner},
  \citenamefont {Strobel},\ and\ \citenamefont
  {Oberthaler}}]{kunkel_simultaneous_2019}%
  \BibitemOpen
  \bibfield  {author} {\bibinfo {author} {\bibfnamefont {P.}~\bibnamefont
  {Kunkel}}, \bibinfo {author} {\bibfnamefont {M.}~\bibnamefont {Pr{\"u}fer}},
  \bibinfo {author} {\bibfnamefont {S.}~\bibnamefont {Lannig}}, \bibinfo
  {author} {\bibfnamefont {R.}~\bibnamefont {{Rosa-Medina}}}, \bibinfo {author}
  {\bibfnamefont {A.}~\bibnamefont {Bonnin}}, \bibinfo {author} {\bibfnamefont
  {M.}~\bibnamefont {G{\"a}rttner}}, \bibinfo {author} {\bibfnamefont
  {H.}~\bibnamefont {Strobel}},\ and\ \bibinfo {author} {\bibfnamefont {M.~K.}\
  \bibnamefont {Oberthaler}},\ }
  \bibfield {title} {\bibinfo {title} \textit{Simultaneous Readout of Noncommuting Collective Spin Observables beyond the Standard Quantum Limit
},\ }
\href
  {https://doi.org/10.1103/PhysRevLett.123.063603} {\bibfield  {journal}
  {\bibinfo  {journal} {Phys. Rev. Lett.}\ }\textbf {\bibinfo {volume} {123}},\
  \bibinfo {pages} {063603} (\bibinfo {year} {2019})}\BibitemShut {NoStop}%
\bibitem [{sup()}]{supp}%
  \BibitemOpen
  \href@noop {} {}\bibinfo {note} {See {{Supplemental Material}} at [{{URL}}
  Will Be Inserted by Publisher] for additional information on the
  experimentral setup, a theoretical description of the prepared squeezed state
  and a theoretical determination of the scale factors, which contains
  Refs.~\cite{anders_momentum_2021, meyer-hoppe_dynamical_2023,
  fang_bess_improving_2018, kawaguchi_spinor_2012, ReinaudiStrong2007,
  luckeTwinSupp2011, lucke_Detecting_2014}.}\BibitemShut {Stop}%
\bibitem [{\citenamefont {{Meyer-Hoppe}}\ \emph {et~al.}(2023)\citenamefont
  {{Meyer-Hoppe}}, \citenamefont {Baron}, \citenamefont {Cassens},
  \citenamefont {Anders}, \citenamefont {Idel}, \citenamefont {Peise},\ and\
  \citenamefont {Klempt}}]{meyer-hoppe_dynamical_2023}%
  \BibitemOpen
  \bibfield  {author} {\bibinfo {author} {\bibfnamefont {B.}~\bibnamefont
  {{Meyer-Hoppe}}}, \bibinfo {author} {\bibfnamefont {M.}~\bibnamefont
  {Baron}}, \bibinfo {author} {\bibfnamefont {C.}~\bibnamefont {Cassens}},
  \bibinfo {author} {\bibfnamefont {F.}~\bibnamefont {Anders}}, \bibinfo
  {author} {\bibfnamefont {A.}~\bibnamefont {Idel}}, \bibinfo {author}
  {\bibfnamefont {J.}~\bibnamefont {Peise}},\ and\ \bibinfo {author}
  {\bibfnamefont {C.}~\bibnamefont {Klempt}},\ }
  \bibfield {title} {\bibinfo {title} \textit{Dynamical low-noise microwave source for cold-atom experiments
},\ }
\href
  {https://doi.org/10.1063/5.0160367} {\bibfield  {journal} {\bibinfo
  {journal} {Rev. Sci. Instrum.}\ }\textbf {\bibinfo {volume} {94}},\ \bibinfo
  {pages} {074705} (\bibinfo {year} {2023})}\BibitemShut {NoStop}%
\bibitem [{\citenamefont {Fang}\ \emph {et~al.}(2018)\citenamefont {Fang},
  \citenamefont {Mielec}, \citenamefont {Savoie}, \citenamefont {Altorio},
  \citenamefont {Landragin},\ and\ \citenamefont
  {Geiger}}]{fang_bess_improving_2018}%
  \BibitemOpen
  \bibfield  {author} {\bibinfo {author} {\bibfnamefont {B.}~\bibnamefont
  {Fang}}, \bibinfo {author} {\bibfnamefont {N.}~\bibnamefont {Mielec}},
  \bibinfo {author} {\bibfnamefont {D.}~\bibnamefont {Savoie}}, \bibinfo
  {author} {\bibfnamefont {M.}~\bibnamefont {Altorio}}, \bibinfo {author}
  {\bibfnamefont {A.}~\bibnamefont {Landragin}},\ and\ \bibinfo {author}
  {\bibfnamefont {R.}~\bibnamefont {Geiger}},\ }
  \bibfield {title} {\bibinfo {title} \textit{Improving the phase response of an atom interferometer by means of temporal pulse shaping
},\ }
\href
  {https://doi.org/10.1088/1367-2630/aaa37c} {\bibfield  {journal} {\bibinfo
  {journal} {New J. Phys.}\ }\textbf {\bibinfo {volume} {20}},\ \bibinfo
  {pages} {023020} (\bibinfo {year} {2018})}\BibitemShut {NoStop}%
\bibitem [{\citenamefont {Hartwig}(2013)}]{hartwig_analyse_2013}%
  \BibitemOpen
  \bibfield  {author} {\bibinfo {author} {\bibfnamefont {J.~M.}\ \bibnamefont
  {Hartwig}},\ }\emph {\bibinfo {title} {{Analyse eines atomaren Gravimeters
  hinsichtlich eines Quantentests des {\"A}quivalenzprinzips}}},\ \href
  {https://doi.org/10.15488/8075} {Ph.D. thesis},\ \bibinfo  {school}
  {Hannover: Gottfried Wilhelm Leibniz Universit{\"a}t Hannover} (\bibinfo
  {year} {2013})\BibitemShut {NoStop}%
\bibitem [{\citenamefont {Riley}\ and\ \citenamefont
  {Howe}(2008)}]{riley_handbook_2008}%
  \BibitemOpen
  \bibfield  {author} {\bibinfo {author} {\bibfnamefont {W.}~\bibnamefont
  {Riley}}\ and\ \bibinfo {author} {\bibfnamefont {D.}~\bibnamefont {Howe}},\
  }\href {https://tsapps.nist.gov/publication/get_pdf.cfm?pub_id=50505} {\emph
  {\bibinfo {title} {Handbook of Frequency Stability Analysis}}}\ (\bibinfo
  {publisher} {Special Publication (NIST SP), National Institute of Standards
  and Technology, Gaithersburg, MD},\ \bibinfo {year} {2008})\BibitemShut
  {NoStop}%
\bibitem [{\citenamefont {Peters}\ \emph {et~al.}(1999)\citenamefont {Peters},
  \citenamefont {Chung},\ and\ \citenamefont {Chu}}]{peters_measurement_1999}%
  \BibitemOpen
  \bibfield  {author} {\bibinfo {author} {\bibfnamefont {A.}~\bibnamefont
  {Peters}}, \bibinfo {author} {\bibfnamefont {K.~Y.}\ \bibnamefont {Chung}},\
  and\ \bibinfo {author} {\bibfnamefont {S.}~\bibnamefont {Chu}},\ }
  \bibfield {title} {\bibinfo {title} \textit{Measurement of gravitational acceleration by dropping atoms
},\ }
\href
  {https://doi.org/10.1038/23655} {\bibfield  {journal} {\bibinfo  {journal}
  {Nature}\ }\textbf {\bibinfo {volume} {400}},\ \bibinfo {pages} {849}
  (\bibinfo {year} {1999})}\BibitemShut {NoStop}%
\bibitem [{\citenamefont {Lezeik}\ \emph {et~al.}(2022)\citenamefont {Lezeik},
  \citenamefont {Tell}, \citenamefont {Zipfel}, \citenamefont {Gupta},
  \citenamefont {Wodey}, \citenamefont {Rasel}, \citenamefont {Schubert},\ and\
  \citenamefont {Schlippert}}]{lezeik_understanding_2022}%
  \BibitemOpen
  \bibfield  {author} {\bibinfo {author} {\bibfnamefont {A.}~\bibnamefont
  {Lezeik}}, \bibinfo {author} {\bibfnamefont {D.}~\bibnamefont {Tell}},
  \bibinfo {author} {\bibfnamefont {K.}~\bibnamefont {Zipfel}}, \bibinfo
  {author} {\bibfnamefont {V.}~\bibnamefont {Gupta}}, \bibinfo {author}
  {\bibfnamefont {{\'E}.}~\bibnamefont {Wodey}}, \bibinfo {author}
  {\bibfnamefont {E.}~\bibnamefont {Rasel}}, \bibinfo {author} {\bibfnamefont
  {C.}~\bibnamefont {Schubert}},\ and\ \bibinfo {author} {\bibfnamefont
  {D.}~\bibnamefont {Schlippert}},\ }
  \bibfield {title} {\bibinfo {title} \textit{Understanding the gravitational and magnetic environment of a very long baseline atom interferometer
},\ }
\href
  {https://doi.org/10.48550/arXiv.2209.08886} {\bibfield  {journal} {\bibinfo
  {journal} {arXiv}\ ,\ \bibinfo {pages} {2209.08886}} (\bibinfo {year}
  {2022})}\BibitemShut {NoStop}%
\bibitem [{\citenamefont {Lotz}\ \emph {et~al.}(2017)\citenamefont {Lotz},
  \citenamefont {Frob{\"o}se}, \citenamefont {Wanner}, \citenamefont
  {Overmeyer},\ and\ \citenamefont {Ertmer}}]{lotz_einsteinelevator_2017}%
  \BibitemOpen
  \bibfield  {author} {\bibinfo {author} {\bibfnamefont {C.}~\bibnamefont
  {Lotz}}, \bibinfo {author} {\bibfnamefont {T.}~\bibnamefont {Frob{\"o}se}},
  \bibinfo {author} {\bibfnamefont {A.}~\bibnamefont {Wanner}}, \bibinfo
  {author} {\bibfnamefont {L.}~\bibnamefont {Overmeyer}},\ and\ \bibinfo
  {author} {\bibfnamefont {W.}~\bibnamefont {Ertmer}},\ }
  \bibfield {title} {\bibinfo {title} \textit{Einstein-Elevator: A New Facility for Research from $\mu$g to 5g
},\ }
\href
  {https://sciendo.com/article/10.2478/gsr-2017-0007} {\bibfield  {journal}
  {\bibinfo  {journal} {Gravitational and Space Research}\ }\textbf {\bibinfo
  {volume} {5}},\ \bibinfo {pages} {11} (\bibinfo {year} {2017})}\BibitemShut
  {NoStop}%
\bibitem [{\citenamefont {Anton}\ \emph {et~al.}(2024)\citenamefont {Anton},
  \citenamefont {Br{\"o}ckel}, \citenamefont {Derr}, \citenamefont {Fieguth},
  \citenamefont {Franzke}, \citenamefont {G{\"a}rtner}, \citenamefont {Giese},
  \citenamefont {Haase}, \citenamefont {Hamann}, \citenamefont {Heidt},
  \citenamefont {Kanthak}, \citenamefont {Klempt}, \citenamefont {Kruse},
  \citenamefont {Krutzik}, \citenamefont {Kubitza}, \citenamefont {Lotz},
  \citenamefont {M{\"u}ller}, \citenamefont {Pahl}, \citenamefont {Rasel},
  \citenamefont {Schiemangk}, \citenamefont {Schleich}, \citenamefont
  {Schwertfeger}, \citenamefont {Wicht},\ and\ \citenamefont
  {W{\"o}rner}}]{anton_INTENTAS_2024}%
  \BibitemOpen
  \bibfield  {author} {\bibinfo {author} {\bibfnamefont {O.}~\bibnamefont
  {Anton}} et al.,\ }
  \bibfield {title} {\bibinfo {title} \textit{INTENTAS - An entanglement-enhanced atomic sensor for microgravity
},\ }
\href
  {https://doi.org/10.48550/arXiv.2409.01051} {\bibfield  {journal} {\bibinfo
  {journal} {arXiv}\ ,\ \bibinfo {pages} {2409.01051}} (\bibinfo {year}
  {2024})}\BibitemShut {NoStop}%
\bibitem [{\citenamefont {Morel}\ \emph {et~al.}(2020)\citenamefont {Morel},
  \citenamefont {Yao}, \citenamefont {Cladé},\ and\ \citenamefont
  {Guellati-Khélifa}}]{Morel_Determination_2020}%
  \BibitemOpen
  \bibfield  {author} {\bibinfo {author} {\bibfnamefont {L.}~\bibnamefont
  {Morel}}, \bibinfo {author} {\bibfnamefont {Z.}~\bibnamefont {Yao}}, \bibinfo
  {author} {\bibfnamefont {P.}~\bibnamefont {Cladé}},\ and\ \bibinfo {author}
  {\bibfnamefont {S.}~\bibnamefont {Guellati-Khélifa}},\ }
  \bibfield {title} {\bibinfo {title} \textit{Determination of the fine-structure constant with an accuracy of 81 parts per trillion
},\ }
\href
  {https://doi.org/10.1038/s41586-020-2964-7} {\bibfield  {journal} {\bibinfo
  {journal} {Nature}\ }\textbf {\bibinfo {volume} {588}},\ \bibinfo {pages}
  {61} (\bibinfo {year} {2020})}\BibitemShut {NoStop}%
\bibitem [{\citenamefont {Fixler}\ \emph {et~al.}(2007)\citenamefont {Fixler},
  \citenamefont {Foster}, \citenamefont {McGuirk},\ and\ \citenamefont
  {Kasevich}}]{Atom_Fixler_2007}%
  \BibitemOpen
  \bibfield  {author} {\bibinfo {author} {\bibfnamefont {J.~B.}\ \bibnamefont
  {Fixler}}, \bibinfo {author} {\bibfnamefont {G.~T.}\ \bibnamefont {Foster}},
  \bibinfo {author} {\bibfnamefont {J.~M.}\ \bibnamefont {McGuirk}},\ and\
  \bibinfo {author} {\bibfnamefont {M.~A.}\ \bibnamefont {Kasevich}},\ }
  \bibfield {title} {\bibinfo {title} \textit{Atom Interferometer Measurement of the Newtonian Constant of Gravity
},\ }
\href
  {https://doi.org/10.1126/science.1135459} {\bibfield  {journal} {\bibinfo
  {journal} {Science}\ }\textbf {\bibinfo {volume} {315}},\ \bibinfo {pages}
  {74} (\bibinfo {year} {2007})}\BibitemShut {NoStop}%
\bibitem [{\citenamefont {Rosi}\ \emph {et~al.}(2014)\citenamefont {Rosi},
  \citenamefont {Sorrentino}, \citenamefont {Cacciapuoti}, \citenamefont
  {Prevedelli},\ and\ \citenamefont {Tino}}]{Rosi_Precision_2014}%
  \BibitemOpen
  \bibfield  {author} {\bibinfo {author} {\bibfnamefont {G.}~\bibnamefont
  {Rosi}}, \bibinfo {author} {\bibfnamefont {F.}~\bibnamefont {Sorrentino}},
  \bibinfo {author} {\bibfnamefont {L.}~\bibnamefont {Cacciapuoti}}, \bibinfo
  {author} {\bibfnamefont {M.}~\bibnamefont {Prevedelli}},\ and\ \bibinfo
  {author} {\bibfnamefont {G.~M.}\ \bibnamefont {Tino}},\ }
  \bibfield {title} {\bibinfo {title} \textit{Precision measurement of the Newtonian gravitational constant using cold atoms
},\ }
\href
  {https://doi.org/10.1038/nature13433} {\bibfield  {journal} {\bibinfo
  {journal} {Nature}\ }\textbf {\bibinfo {volume} {510}},\ \bibinfo {pages}
  {518} (\bibinfo {year} {2014})}\BibitemShut {NoStop}%
\bibitem [{\citenamefont {Schrinski}\ \emph {et~al.}(2023)\citenamefont
  {Schrinski}, \citenamefont {Haslinger}, \citenamefont {Schmiedmayer},
  \citenamefont {Hornberger},\ and\ \citenamefont
  {Nimmrichter}}]{Schrinski_testing_2023}%
  \BibitemOpen
  \bibfield  {author} {\bibinfo {author} {\bibfnamefont {B.}~\bibnamefont
  {Schrinski}}, \bibinfo {author} {\bibfnamefont {P.}~\bibnamefont
  {Haslinger}}, \bibinfo {author} {\bibfnamefont {J.}~\bibnamefont
  {Schmiedmayer}}, \bibinfo {author} {\bibfnamefont {K.}~\bibnamefont
  {Hornberger}},\ and\ \bibinfo {author} {\bibfnamefont {S.}~\bibnamefont
  {Nimmrichter}},\ }
  \bibfield {title} {\bibinfo {title} \textit{Testing collapse models with Bose-Einstein-condensate interferometry
},\ }
\href {https://doi.org/10.1103/PhysRevA.107.043320}
  {\bibfield  {journal} {\bibinfo  {journal} {Phys. Rev. A}\ }\textbf {\bibinfo
  {volume} {107}},\ \bibinfo {pages} {043320} (\bibinfo {year}
  {2023})}\BibitemShut {NoStop}%
\bibitem [{\citenamefont {Cassess}\ \emph {et~al.}(2025)\citenamefont
  {Cassess}, \citenamefont {Meyer-Hoppe}, \citenamefont {Rasel},\ and\
  \citenamefont {Klempt}}]{CassensRepository2025}%
  \BibitemOpen
  \bibfield  {author} {\bibinfo {author} {\bibfnamefont {C.}~\bibnamefont
  {Cassess}}, \bibinfo {author} {\bibfnamefont {B.}~\bibnamefont
  {Meyer-Hoppe}}, \bibinfo {author} {\bibfnamefont {E.}~\bibnamefont {Rasel}},\
  and\ \bibinfo {author} {\bibfnamefont {C.}~\bibnamefont {Klempt}},\ }
  {\bibinfo {title} {Data supporting
  publication 'An entanglement-enhanced atomic gravimeter'}} (\bibinfo {year}
  {2025})
  \bibfield {doi} \href
  {https://doi.org/10.25835/VCCCUUHF} {\bibinfo {doi} \textit{10.25835/VCCCUUHF} }\BibitemShut {NoStop}%
\bibitem [{\citenamefont {Reinaudi}\ \emph {et~al.}(2007)\citenamefont
  {Reinaudi}, \citenamefont {Lahaye}, \citenamefont {Wang},\ and\ \citenamefont
  {Gu\'{e}ry-Odelin}}]{ReinaudiStrong2007}%
  \BibitemOpen
  \bibfield  {author} {\bibinfo {author} {\bibfnamefont {G.}~\bibnamefont
  {Reinaudi}}, \bibinfo {author} {\bibfnamefont {T.}~\bibnamefont {Lahaye}},
  \bibinfo {author} {\bibfnamefont {Z.}~\bibnamefont {Wang}},\ and\ \bibinfo
  {author} {\bibfnamefont {D.}~\bibnamefont {Gu\'{e}ry-Odelin}},\ }
  \bibfield {title} {\bibinfo {title} \textit{Strong saturation absorption imaging of dense clouds of ultracold atoms
},\ }
\href
  {https://doi.org/10.1364/OL.32.003143} {\bibfield  {journal} {\bibinfo
  {journal} {Opt. Lett.}\ }\textbf {\bibinfo {volume} {32}},\ \bibinfo {pages}
  {3143} (\bibinfo {year} {2007})}\BibitemShut {NoStop}%
\bibitem [{\citenamefont {L{\"u}cke}\ \emph {et~al.}(2011)\citenamefont
  {L{\"u}cke}, \citenamefont {Scherer}, \citenamefont {Kruse}, \citenamefont
  {Pezz{\'e}}, \citenamefont {Deuretzbacher}, \citenamefont {Hyllus},
  \citenamefont {Topic}, \citenamefont {Peise}, \citenamefont {Ertmer},
  \citenamefont {Arlt}, \citenamefont {Santos}, \citenamefont {Smerzi},\ and\
  \citenamefont {Klempt}}]{luckeTwinSupp2011}%
  \BibitemOpen
  \bibfield  {author} {\bibinfo {author} {\bibfnamefont {B.}~\bibnamefont
  {L{\"u}cke}}, \bibinfo {author} {\bibfnamefont {M.}~\bibnamefont {Scherer}},
  \bibinfo {author} {\bibfnamefont {J.}~\bibnamefont {Kruse}}, \bibinfo
  {author} {\bibfnamefont {L.}~\bibnamefont {Pezz{\'e}}}, \bibinfo {author}
  {\bibfnamefont {F.}~\bibnamefont {Deuretzbacher}}, \bibinfo {author}
  {\bibfnamefont {P.}~\bibnamefont {Hyllus}}, \bibinfo {author} {\bibfnamefont
  {O.}~\bibnamefont {Topic}}, \bibinfo {author} {\bibfnamefont
  {J.}~\bibnamefont {Peise}}, \bibinfo {author} {\bibfnamefont
  {W.}~\bibnamefont {Ertmer}}, \bibinfo {author} {\bibfnamefont
  {J.}~\bibnamefont {Arlt}}, \bibinfo {author} {\bibfnamefont {L.}~\bibnamefont
  {Santos}}, \bibinfo {author} {\bibfnamefont {A.}~\bibnamefont {Smerzi}},\
  and\ \bibinfo {author} {\bibfnamefont {C.}~\bibnamefont {Klempt}},\ }
  \bibfield {title} {\bibinfo {title} \textit{Twin Matter Waves for Interferometry Beyond the Classical Limit
},\ }\href
  {https://doi.org/10.1126/science.1208798} {\bibfield  {journal} {\bibinfo
  {journal} {Science}\ }\textbf {\bibinfo {volume} {334}},\ \bibinfo {pages}
  {773} (\bibinfo {year} {2011})}\BibitemShut {NoStop}%
\bibitem [{\citenamefont {L{\"u}cke}\ \emph {et~al.}(2014)\citenamefont
  {L{\"u}cke}, \citenamefont {Peise}, \citenamefont {Vitagliano}, \citenamefont
  {Arlt}, \citenamefont {Santos}, \citenamefont {T{\'o}th},\ and\ \citenamefont
  {Klempt}}]{lucke_Detecting_2014}%
  \BibitemOpen
  \bibfield  {author} {\bibinfo {author} {\bibfnamefont {B.}~\bibnamefont
  {L{\"u}cke}}, \bibinfo {author} {\bibfnamefont {J.}~\bibnamefont {Peise}},
  \bibinfo {author} {\bibfnamefont {G.}~\bibnamefont {Vitagliano}}, \bibinfo
  {author} {\bibfnamefont {J.}~\bibnamefont {Arlt}}, \bibinfo {author}
  {\bibfnamefont {L.}~\bibnamefont {Santos}}, \bibinfo {author} {\bibfnamefont
  {G.}~\bibnamefont {T{\'o}th}},\ and\ \bibinfo {author} {\bibfnamefont
  {C.}~\bibnamefont {Klempt}},\ }
  \bibfield {title} {\bibinfo {title} \textit{Detecting Multiparticle Entanglement of Dicke States
},\ }
\href
  {https://doi.org/10.1103/PhysRevLett.112.155304} {\bibfield  {journal}
  {\bibinfo  {journal} {Phys. Rev. Lett.}\ }\textbf {\bibinfo {volume} {112}},\
  \bibinfo {pages} {155304} (\bibinfo {year} {2014})}\BibitemShut {NoStop}%
\bibitem [{\citenamefont {Kawaguchi}\ and\ \citenamefont
  {Ueda}(2012)}]{kawaguchi_spinor_2012}%
  \BibitemOpen
  \bibfield  {author} {\bibinfo {author} {\bibfnamefont {Y.}~\bibnamefont
  {Kawaguchi}}\ and\ \bibinfo {author} {\bibfnamefont {M.}~\bibnamefont
  {Ueda}},\ }
  \bibfield {title} {\bibinfo {title} \textit{Spinor Bose–Einstein condensates
},\ }
\href {https://doi.org/10.1016/j.physrep.2012.07.005} {\bibfield
  {journal} {\bibinfo  {journal} {Physics Reports}\ }\textbf {\bibinfo {volume}
  {520}},\ \bibinfo {pages} {253} (\bibinfo {year} {2012})}\BibitemShut
  {NoStop}%
  

\end{thebibliography}

%

\end{document}

\title{An entanglement-enhanced atomic gravimeter (Supplemental Material)}

\author{Christophe Cassens$^{1}$}\email[Author to whom correspondence should be addressed: ]{Christophe.Cassens@dlr.de}
\author{Bernd Meyer-Hoppe$^{1}$}
\author{Ernst Rasel$^{1}$}
\author{Carsten Klempt$^{1,2}$}
%Add potential coauthors: (D. Schlippert, C. Schubert abwarten ob Hilfe bei Systematik), F. Anders, T. Sanchez

\affiliation{$^1$Leibniz Universit\"at Hannover, Institut für Quantenoptik, Welfengarten 1, D-30167 Hannover, Germany  \\
$^2$Deutsches Zentrum f\"ur Luft- und Raumfahrt e.V. (DLR), Institut f\"ur Satellitengeod\"asie und Inertialsensorik, Callinstraße 30b, D-30167 Hannover, Germany
}

\date{\today}
\maketitle
\setcounter{figure}{3} 
\section*{Supplementary Material}

\subsection*{Raman laser setup}
\begin{figure*}[ht]
\centering
  \includegraphics[width=0.8\textwidth]{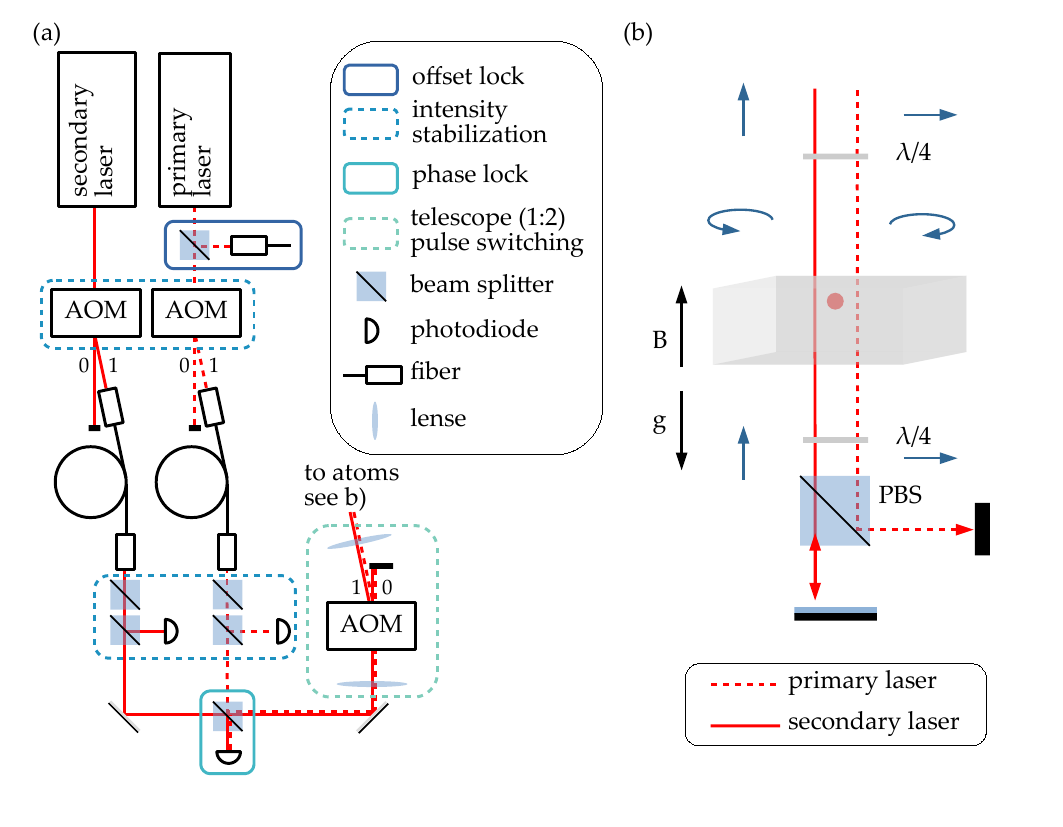}
    \caption{
    Setup of the Raman optics and beam configuration.
    (a) Optical setup for intensity and phase stabilization as well as pulse forming.
    The optical frequencies are provided by two ECDLs. 
    The primary laser's frequency is stabilized with respect to the cooling laser with an offset of $-2\,$GHz.
    %To this end part of the primary laser beam is coupled into a fiber leading to the cooling laser.
    The intensities of the secondary and primary lasers are stabilized behind a mode-cleaning fiber of $1\,$m length.
    The intensities are measured by photodiodes and used as an input to control the laser intensities by PID controllers.
    The intensities are adjusted by actuating the rf power that drives the acousto-optical modulators (AOMs) before the fibers.
    The beams of the secondary and primary lasers are overlapped on a polarizing beam splitter (PBS).
    One output leads to a photodiode with high bandwidth measuring the beat signal for the phase stabilization of the secondary laser.
    The other output is expanded to $3\,$mm beam diameter by a telescope.
    In the focus of the first lens, an AOM is placed for simultaneous pulse forming and switching of both beams.
    (b) Beam configuration for momentum transfer. 
    The expanded, overlapped and perpendicularly linearly polarized (blue arrows) beams of the secondary and primary lasers are altered into counter rotating circular polarization before entering the vacuum chamber and interacting with the BEC.
    Upon leaving the chamber the polarizations are changed back and the primary laser beam is coupled out by a PBS.
    The secondary laser beam is retroreflected into the chamber by a mirror perpendicular to Earth's gravitational field.}\label{fig:ramanSetup}
\end{figure*}

The momentum mode of the BEC is changed via Raman transitions of two counter-propagating laser beams. 
\blue{Figure \ref{fig:ramanSetup}(a) shows the optical setup involved in providing the phase and intensity stabilized optical frequencies.}
The laser beams are generated by two external-cavity diode lasers (ECDLs). 
The primary laser is frequency-stabilized with respect to the D$_2$-transition with an offset of $-2\,$GHz.
The secondary laser is phase-locked to the primary at a dynamically changeable frequency difference of about $6.83\,$GHz. 
The spatial modes of each laser is cleaned by one meter of single-mode optical fiber.
The intensities are individually stabilized using a photodiode after the fibre and an acousto-optical modulator (AOM) before the fibre.
After the fibre, the two linearly polarized beams are coupled on a polarizing beam splitter.
One output port leads to a high-bandwidth photodiode to generate the signal for the phase lock.
The other port is coupled into a switching AOM which is controlled by a direct digital synthesizer of our mw system \cite{meyer-hoppe_dynamical_2023}. 
The detuning with respect to the cooling laser transition is reduced to $-1.72\,$GHz due to the added frequency from the AOMs.
The beams are changed into counter-rotating circular polarization before entering the experimental chamber \blue{(Fig. \ref{fig:ramanSetup}(b))}.
After leaving the experimental chamber, the beams are again converted to perpendicular linear polarizations, such that the first laser can be extracted by a polarizing beam splitter.
The second laser is retroreflected towards the incoming beam, changing its circular polarization handedness at the position of the atoms \cite{anders_momentum_2021}.
This leads to only one combination of beams driving Raman transitions since all other are suppressed by choice of polarization.

\subsection*{Characterization of the Raman coupling}
The presence of light fields shifts atomic energy levels inversely proportional to the detuning from the respective transition (AC Stark shift). 
In our setup the states $|F=2,m=0\rangle$ and $|F=1,m=0\rangle$ are shifted proportional to the intensity of the primary and secondary laser, respectively.
Due to the employed interferometer sequence, common light shifts do not contribute to the interferometer's phase.
Moreover, the differential AC Stark shift component can be suppressed by adjusting the relative intensity between primary and secondary laser.
The optimal intensity ratio is experimentally determined by reducing the light-shift-induced phase accumulation during a mw Ramsey sequence due to a Raman pulse with $1\,$MHz detuning. 
To avoid being on the wrong fringe we start with short pulse duration which is only increased once the mid-fringe point is reached.  

The noise contribution to the gravimeter signal by the AC Stark shift is estimated from the increased variance of a mw Mach-Zehnder interferometer.
To this end, a $2\pi\times1$\,MHz detuned and $60\,$\textmu s Blackman-shaped Raman pulse is applied before and after the mw $\pi$-pulse in the mw interferometer sequence and the resulting noise level is compared to a sequence without Raman pulses.
With light shift, the noise level is $1.3^{+0.4}_{-0.4}\,$dB and without light shifts the interferometer signal features noise $0.5^{+0.4}_{-0.5}\,$dB above the SQL.
Considering the technical noise level due to intensity fluctuations from these measurements, one would expect a degradation of the spin squeezing due to light shifts to $-2.6^{+1.3}_{-1.9}\,$dB.
The AC Stark shift is thus the main noise source giving rise to the apparent reduction of the squeezing in the interferometer sequence.

\begin{figure}[ht]
\centering
  \includegraphics[width=\columnwidth]{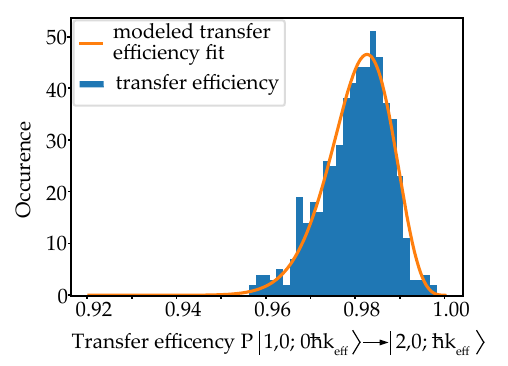}
    \caption{
    Raman $\pi$-pulse efficency. Transfer efficiency from $|F=1, m=0; p=0\rangle$ to $|F=2, m=0; p=\hbar k \rangle$ of a $64.8\,$\textmu{}s Blackman-shaped Raman pulse. 
    The mean transfer efficiency of 510 consecutive measurements is $98.1(7)\,\%$.
    Also shown is a fitted model of the transfer efficiency corresponding to an AC-Stark-shift induced detuning of $2\pi \times 2.5\,$kHz with fluctuations of $2\pi \times 0.5\,$kHz.}\label{fig:ramanPiPulseEfficiency}
\end{figure}

For a $\pi$-pulse, the transfer efficiency is measured to be $98.1(7)\,\%$ (Fig.~\ref{fig:ramanPiPulseEfficiency}).
In particular due to the chosen interferometer geometry, where parasitic paths are detached from the main path, this efficiency does not deteriorate the retrieved signal.
%Also shown is a modeled probability of transfer efficiency under the influence of %frequency deviation from resonance

%\begin{align*}
    %p(P) = \frac{1}{\Delta_\text{fluc}\sqrt{2\pi}}e^{-\frac{1}{2}\left(\frac{\omega_0%\sqrt{1-P}-\Delta_0}{\Delta_\text{fluc}}\right)^2}
%\end{align*}
%
%assuming a constant detuning of $\Delta_0 = 2\pi \times 2.5\,$kHz and fluctuation of %$\Delta_\text{fluc}=2\pi\times0.5\,$kHz, Rabi frequency
%$\Omega_0$ and transfer efficency $P$.

From weighted out-of-loop phase noise measurements we estimate a phase noise of $\sigma_\phi=1.2\,$mrad for $\tau_\pi = 25\,$µs, $T=455\,$µs and $T_\text{sep}=50\,$µs.
For the employed atom number of $N=6000$, this corresponds to a noise contribution in $J_\text{z}$ of $\Delta J_\text{z}=4$ or $-20\,$dB below the SQL.
Phase noise is therefore a negligible noise contribution in our experiments.

\subsection*{Detection}
\blue{
The experimental sequence is completed by a detection of the number of atoms in the interferometer's output ports.
To this end, the atoms in the level $|1,0\rangle$ are transferred to the level $|2,1\rangle$.
%Due to the resonance frequency of the transition $\vert 2,0\rangle \rightarrow \vert 1,1\rangle$ being only $2\pi\times2.5\,$kHz detuned this transition is also partly driven.
This coupling pulse also transfers a part of the atoms in $\vert 2,0\rangle$ to $\vert 1,1\rangle$, which we further move to $\vert 2,2\rangle$ for its detection.
%A subsequent mw pulse transfers those atoms to $\vert 2,2 \rangle$.
%In summary the original output port $\vert 1,0 \rangle$ has been mapped to $\vert 2,1 \rangle$
%and the original output port $\vert 2,0 \rangle$ has been mapped to $\vert 2,0 \rangle$ and $\vert 2,2 \rangle$.
During $4.2$~ms of free fall, an inhomogeneous magnetic field separates the atoms according to their Zeeman-state.
The number of atoms in the three clouds is recorded by absorption detection, where the atoms are illuminated by a resonant laser beam with an intensity of $2.5\,\frac{\text{mW}}{\text{cm}^2}$.
The shadow of the atoms is imaged onto a high-efficiency CCD camera and evaluated to yield an absolute number per pixel~\cite{ReinaudiStrong2007, luckeTwinSupp2011}.
The final result is obtained by a summation over three separate spatial regions on the CCD chip.}

\blue{We perform an independent test of the number calibration by measuring shot noise fluctuations.
A BEC in the level $\vert 1, 0 \rangle$ is put into a coherent superposition of the clock states by a resonant mw pulse and the variance of $J_\text{z}$ is measured.
To exclude the noise contribution due to the mw setup, the variance is determined for three different pulse lengths ($\frac{\pi}{2}$, $\frac{5\pi}{2}$, and $\frac{9\pi}{2}$) to fit and subtract the microwave-induced noise component.
We further subtract the detection noise, resulting from shot noise of the photoelectrons on the camera pixels, which we obtain from empty images, Var($J_{\text{z, }N=0}) = 213$.
Additional number-dependent noise remains negligible~\cite{lucke_Detecting_2014}.
The resulting fluctuations Var($J_\text{z}$) are $3.8\,\% \pm 5.4\,\%$ above shot noise which is compatible with our number calibration.}

%\begin{figure}[t!]
%\centering
%  \includegraphics[width=\columnwidth]{Gravimeter/figures/figures_actually_in_paper/figure_5_detectionCalibration.pdf}
%    \caption{ Independent check of atom number counting fidelity. A BEC of about $14000$ atoms prepared in $\vert 1, 0 \rangle$ is put into a coherent superposition on the clock transition by a mw $\frac{\pi}{2}$-, $\frac{5\pi}{2}$- and $\frac{9\pi}{2}$-pulse. 
%    After applying the magnetic field gradient each detected cloud consists of at most $7000$ atoms. 
%    The variances for each pulse duration are fitted to obtain the noise contribution $a$ which is free of mw pulse dependent noise.
%    The black line is the SQL according to the estimated number of atoms. The pulse duration is given in multiples of the $\frac{\pi}{2}$-pulse duration $\tau_\frac{\pi}{2}$.
%  }\label{fig:detectionCalibration}
%\end{figure}

\subsection*{Single-mode squeezing}
The Hamiltonian describing spin-changing collisions is given by~\cite{kawaguchi_spinor_2012}
\begin{align}\nonumber
    \hat{H}=&q\left( \hat{N}_{+1} + \hat{N}_{-1} \right) - \frac{\Omega}{N} \left[ \left( \hat{N}_0-\frac{1}{2} \right)\left( \hat{N}_{+1} + \hat{N}_{-1} \right) \right. \\
    &+ \left.  \hat{a}_{0}^{\dagger}\hat{a}_{0}^{\dagger}\hat{a}_{+1}\hat{a}_{-1} + \hat{a}_{+1}^{\dagger}\hat{a}_{-1}^{\dagger}\hat{a}_{0}\hat{a}_{0} \right]
\end{align}
\blue{with spin interaction strength
\begin{equation}\nonumber
    \Omega=-g_2 \int d^3 r |\Psi(\vec{r})|^4 \text{,}
\end{equation}
$g_2=4\pi\hbar^2(a_2^\text{sc}-a_0^\text{sc})/(3\,m_\text{Rb-87})$ with scattering lengths $a_i^\text{sc}$ and atomic mass $m_\text{Rb-87}$,} quadratic Zeeman energy $q$ and the bosonic annihilation (creation) operators $a_m^{(\dagger)}$ for level $|1,m\rangle$.
Number operators are defined as $\hat{N}_m=\hat{a}_m^{\dagger}\hat{a}_m$.
In the undepleted pump approximation, $N_0\approx N \gg 1$, this results in 
\begin{equation}\nonumber
    \hat{H}= \left( q - \Omega  \right) \left( \hat{N}_{+1} + \hat{N}_{-1} \right) - \Omega \left( \hat{a}_{+1}\hat{a}_{-1} + \hat{a}_{+1}^{\dagger}\hat{a}_{-1}^{\dagger} \right) \text{.}
\end{equation}
When the quadratic Zeeman energy is now tuned to $q=\Omega$ by mw dressing, we are left with the two-mode squeezing Hamiltonian
\begin{equation}\nonumber
    \hat{H}= -\Omega \left( \hat{a}_{+1}\hat{a}_{-1} + \hat{a}_{+1}^{\dagger}\hat{a}_{-1}^{\dagger} \right) 
\end{equation}
that generates entangled pairs of atoms in the $+1$ and $-1$ levels.
Introducing the symmetric and antisymmetric operators
\begin{equation}\nonumber
    \hat{a}_s=\frac{1}{\sqrt{2}} \left( \hat{a}_{+1}+\hat{a}_{-1} \right)
\end{equation}
and
\begin{equation}\nonumber
    \hat{a}_a= \frac{1}{\sqrt{2}} \left( \hat{a}_{+1}-\hat{a}_{-1} \right)  \text{,}
\end{equation}
we can rewrite the Hamiltonian as
\begin{align}
    \hat{H}&=\frac{\Omega}{2} \left( \hat{a}_{a}\hat{a}_{a} + \hat{a}_{a}^{\dagger}\hat{a}_{a}^{\dagger}\right) -\frac{\Omega}{2} \left( \hat{a}_{s}\hat{a}_{s} + \hat{a}_{s}^{\dagger}\hat{a}_{s}^{\dagger} \right)  \\ \nonumber
    &=\hat{H}_a - \hat{H}_s \text{.}
\end{align}
This presents a combination of two single-mode squeezing operators for the symmetric ($\hat{H}_s$) and the antisymmetric ($\hat{H}_a$) superposition.
By applying a rf pulse after the squeezing generation that only transfers the symmetric superposition from $|1,\pm 1\rangle$, we obtain a single-mode squeezed state in $|1,0\rangle$ (Fig.~\ref{fig:sequenceScheme} (a) and (b)).

\blue{The symmetric and antisymmetric single-mode squeezing Hamiltonians derived above correspond to the quadrature squeezing operators well established in quantum optics \cite{kruse_improvement_2016}.
Therefore, we can estimate the amount of atoms per mode as follows
\begin{align*}
\langle N_{s} \rangle = \langle N_{a} \rangle = \sinh^2(r)
\end{align*}
with $r$ being the squeeze parameter.
In analogy to quantum optics it is known that the fluctuation of the antisqueezed quadrature, in our case $\text{Var}(J_\text{z})$ at phase $\varphi_\text{opt}-\tfrac{\pi}{2}$ (see Fig.~\ref{fig:sequenceScheme} (b)), is increased by a factor of $e^{2r}$ compared to the coherent state.
From the measured increase of the fluctuations by a factor of $9.9\,\text{dB} \,\hat{=}\, 9.8$ we obtain a mean population of the symmetric mode of $\langle N_{s} \rangle = \sinh^2(r) = 1.9$. 
The antisqueezed quadrature is chosen in contrast to the squeezed quadrature, since its variance is dominated by quantum fluctuations instead of technical noise.}

\subsection*{Theoretical determination of the scale factor}
In order to derive the local acceleration from the recorded normalized population in $|2,0\rangle$, the scale factor $S(T)$ (Eq.~\eqref{eq:gravimeterPhase}) has to be determined. 

The Blackman pulse shape \cite{meyer-hoppe_dynamical_2023} has to be taken into account \cite{fang_bess_improving_2018},
\begin{align*}
    \Omega_\text{bm}(t, t_0) = \Omega_0\left(0.42 - \frac{1}{2}\text{cos}\left(2\pi\tilde{t}\right)+0.08 \text{cos}\left(4\pi\tilde{t}\right)\right),
\end{align*}
where $\Omega_0$ is the Rabi frequency corresponding to the maximal intensity and $\tilde{t} = t - t_0$ is the time passing since the beginning of the pulse $t_0$.

For the Blackman pulse shape, the sensitivity function $g_\text{bm}(t)$ is given by
\begin{align*}
    g_\text{bm}(t) = \text{sin}\left(\int^t_{t_0}\Omega_\text{bm}(t' ,t_0)\text{d}t' \right).
\end{align*}

The sensitivity function  of the whole gravimeter sequence $g_\text{grav}(t)$ is then given by
\begin{widetext}
\begin{align}
g_\text{grav}(t) =
\begin{cases}
    g_\text{bm}(t,t_0),\quad t_0 <  t < \tau_\text{bm}, \\
    1,\quad \tau_\text{bm} < t < \tau_\text{bm} + T_\text{R}, \\
    1 - g_\text{bm}(t, t_0 + \tau_\text{bm}+T_\text{R}),\quad \tau_\text{bm}+T_\text{R} < t < 2\tau_\text{bm}+T_\text{R},\\
    0,\quad 2\tau_\text{bm}+T_\text{R} < t < 2\tau_\text{bm}+T_\text{R}+T,\\
    - g_\text{bm}(t, t_0 + 2\tau_\text{bm}+T_\text{R}+T),\quad 2\tau_\text{bm}+T_\text{R}+T < t < 3\tau_\text{bm}+T_\text{R}+T,\\
    -1,\quad 3\tau_\text{bm}+T_\text{R}+T < t < 3\tau_\text{bm}+2T_\text{R}+T,\\
    -1 + g_\text{bm}(t, t_0 + 3\tau_\text{bm}+2T_\text{R}),\quad 3\tau_\text{bm}+2T_\text{R}+T < t < 4\tau_\text{bm}+2T_\text{R}+T,\\
    0,\quad  \text{otherwise.}
    \end{cases}
\end{align}
\end{widetext}

The scale factor $S(T)$ for given Blackman pulse duration $\tau_\text{bm}=60\,$µs and delocalization time $T_\text{R}=77\,$µs evaluates to
\begin{align}
    S(T) = k_\text{eff}\int_{t_0}^{4\tau_\text{bm}+2T_\text{R}+T}g_\text{grav}(t_0, t)\text{d}t.
\end{align}

From this we get $S(T=455\,\text{\textmu s}) = 1.4290\,\frac{\text{s}}{\text{m}^2}$ and $S(T=155\,\text{\textmu s}) = 0.7707\,\frac{\text{s}}{\text{m}^2}$ which differs from the scale factors measured in Fig.~\ref{fig:chirprate}(b) by only $0.7\,\%$ and $0.5\,\%$.

For a small phase range around the mid-fringe position, the respective scale factors thus approximate the measured slopes $\partial z/\partial g$ very well and the measured deviation remains small in comparison to the statistical fluctuations.